\documentclass[12pt]{article}
\usepackage{amsmath}
\usepackage{times}
\usepackage{graphicx}
\usepackage{color}
\usepackage{multirow}
\usepackage[comma,numbers]{natbib} 
\usepackage{rotating}
\usepackage{bbm}
\usepackage{latexsym}
\usepackage{hyperref}
\usepackage{floatrow}

\usepackage[T1]{fontenc}
\usepackage[scaled]{helvet}

\newcommand{\captionfonts}{\normalsize}

\makeatletter  
\long\def\@makecaption#1#2{%
  \vskip\abovecaptionskip
  \sbox\@tempboxa{{\captionfonts #1: #2}}%
  \ifdim \wd\@tempboxa >\hsize
    {\captionfonts #1: #2\par}
  \else
    \hbox to\hsize{\hfil\box\@tempboxa\hfil}%
  \fi
  \vskip\belowcaptionskip}
\makeatother   

\begin{document}
\hspace{13.9cm}

\vspace{50pt}
\noindent 
{\LARGE Capturing the Dynamical Repertoire of Single Neurons with Generalized Linear Models} 

\ \\
{\large {\bf Alison I. Weber}$^{\text 1}$ {\bf \& Jonathan W. Pillow}$^{\text 2, \text 3}$}\\
{$^{\text 1}$Graduate Program in Neuroscience, University of Washington, Seattle, WA, USA.}\\
{$^{\text 2}$Princeton Neuroscience Institute, Princeton University, Princeton, NJ, USA.}\\
{$^{\text 3}$Dept. of Psychology, Princeton University, Princeton, NJ, USA.}\\
%


\thispagestyle{empty}

\begin{center} {\bf Abstract} \end{center} 
A key problem in computational neuroscience is to find simple,
tractable models that are nevertheless flexible enough to capture the
response properties of real neurons. Here we examine the capabilities
of recurrent point process models known as Poisson generalized linear
models (GLMs). These models are defined by a set of linear filters, a
point nonlinearity, and conditionally Poisson spiking.  They have
desirable statistical properties for fitting and have been widely used to analyze
spike trains from electrophysiological recordings. However, the
dynamical repertoire of GLMs has not been systematically compared to
that of real neurons.  Here we show that GLMs can reproduce a comprehensive suite of
canonical neural response behaviors, including tonic and phasic
spiking, bursting, spike rate adaptation, type I and type II
excitation, and two forms of bistability. GLMs can also capture
stimulus-dependent changes in spike timing precision and reliability
that mimic those observed in real neurons, and can exhibit varying
degrees of stochasticity, from virtually deterministic responses to
greater-than-Poisson variability.  These results show that Poisson
GLMs can exhibit a wide range of dynamic spiking behaviors found in
real neurons, making them well suited for qualitative dynamical as
well as quantitative statistical studies of single-neuron and
population response properties.

\newpage


\section{Introduction}
Understanding the dynamical and computational properties of neurons is
a fundamental challenge in cellular and systems neuroscience. A wide
variety of single-neuron models have been proposed to account for
neural response properties. These models can be arranged along a
complexity axis ranging from detailed, interpretable, biophysically
accurate models to simple, tractable, reduced functional
models. Detailed Hodgkin-Huxley style models, which sit at one end of
this continuum, provide a biophysically detailed account of the
conductances, currents, and channel kinetics governing neural response
properties \citep{HodgkinHuxley52}. These models can account for the
vast dynamical repertoire of real neurons, but they are often unwieldy
for theoretical analyses of neural coding and computation. This
motivates the need for simplified models of neural spike responses
that are tractable enough for mathematical, computational, and
statistical analyses.

A variety of simplified dynamical models have been proposed to
serve the need for mathematically tractable models, including the
integrate-and-fire model, Fitzhugh-Nagumo, Morris-Lecar, and
Izhikevich models
\citep{Fitzhugh61,Nagumo62,Morris81,Izhikevich03,Brette05}. Generally, these
models aim to reduce the biophysically detailed descriptions of
realistic neurons to systems of differential equations with fewer
variables and/or simplified dynamics. The one-dimensional
integrate-and-fire model is arguably the simplest of these, and the
simplest to analyze mathematically, but it fails to capture many of the
response properties of real neurons. The two-dimensional Izhikevich
model, by contrast, was specifically formulated to retain the rich
dynamical repertoire of more complex, biophysically realistic models
\citep{Izhikevich04}.

An alternative to a mathematical notion of simplicity is the
statistical property of being tractable for fitting from intracellular
or extracellular physiological recordings.  One well-known statistical
model that satisfies this desideratum is the recurrent
linear-nonlinear Poisson model, commonly referred to in the
neuroscience literature as the {\it generalized linear model} (GLM)
\citep{Truccolo05,Pillow08}.  GLMs are closely related to generalized
integrate-and-fire models such as the spike-response model, which has
linear dynamics but incorporates spike-dependent feedback to capture
the nonlinear effects of spiking on neural membrane potential and
subsequent spike generation
\citep{Gerstner01,Keat01,Jolivet03,Pillow05,Gerstner96}. In fact, a variant of the spike response model that incorporates noise into the spike threshold is mathematically equivalent to the models we study here \citep{Gerstner92,Gerstner95,Jolivet06,Gerstner14}. GLMs are popular
for characterizing neural responses in reverse-correlation or
white-noise experiments, due to the tractability of likelihood-based
fitting methods.  Recent work has shown that GLMs can capture the
detailed statistics of spiking in single and multi-neuron recordings
from a variety of brain areas
\citep{Pillow08,Babadi2010,Calabrese11,Weber12,Mensi12,Pozzorini13}.

While several studies have shown that GLMs can successfully recapitulate various response properties of biological or simulated neurons, here, we provide a more systematic study of the dynamical repertoire of the GLM .  We study
this issue by fitting GLMs to data from simulated neurons exhibiting a
number of complex response properties. We show that GLMs can reproduce
a remarkably rich set of dynamical behaviors, including tonic and
phasic spiking, bursting, spike rate adaptation, type I and type II
excitation, and two different forms of bistability. Furthermore, GLMs
can exhibit stimulus-dependent degrees of spike timing precision and
reliability \citep{Mainen95}, and mimic a recently reported form of
greater-than-Poisson variability \citep{Goris14}.

\section{Models of dynamical behaviors}

\subsection{Izhikevich model}

First, we will examine whether generalized linear models can reproduce
a suite of canonical spiking behaviors exhibited by the well-known
Izhikevich model \citep{Izhikevich03,Izhikevich04}. The Izhikevich
model is a biophysically-inspired model of intracellular membrane
potential defined by a two-variable system of ordinary differential
equations governing membrane potential $v(t)$ and a recovery variable
$u(t)$:
\begin{eqnarray}
\dot v = 0.04v^2+5v+140-u+I(t) \\
\dot u = a(bv-u)
\end{eqnarray}
with spiking and voltage-reset governed by the boundary condition:
\begin{equation}
\text{if  } v(t) \geq 30,\;  \text{``spike''  and set} 
\left\{
        \begin{array}{ll}
            v(t^+) = c \\
            u(t^+) = u(t)+d,
        \end{array}
    \right.
\end{equation}
where $I(t)$ is injected current, $t^+$ denotes the next time step
after $t$, and parameters $(a,b,c,d)$ determine the model's
dynamics. Different settings of these parameters lead to qualitatively
different spiking behaviors, as shown in \citep{Izhikevich04}. We focus
on this model because of its demonstrated ability to produce a wide
range of response properties exhibited by real neurons.  (See Table 1
for parameter values used in this study and Methods for simulation
details.)

\subsection{Generalized linear model (GLM)}

The GLM is a regression model typically used to characterize the
relationship between external or internal covariates and a set of
recorded spike trains.  In systems neuroscience, the label ``GLM''
often refers to an autoregressive point process model, a model in
which linear functions of stimulus and spike history are nonlinearly
transformed to produce the spike rate or {\it conditional intensity}
of a Poisson process \citep{Truccolo05,Pillow08}.

\begin{figure}[h!]
\includegraphics[width=\textwidth]{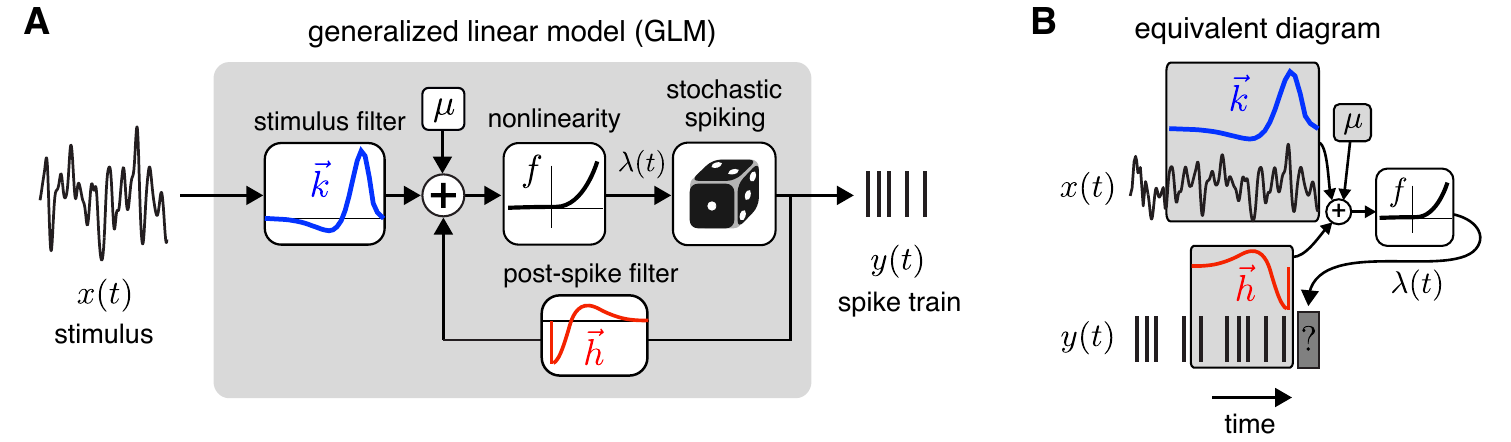} 
\caption{{\bf Schematic of the generalized linear model.}
A: The stimulus filter $\vec k$ operates linearly on the stimulus
  $x(t)$, is combined with input from the post-spike filter $\vec h$
  and mean input level $\mu$.  This combined linear signal passes
  through a point nonlinearity $f(\cdot)$, whose output drives spiking
  via a conditionally Poisson process. B: An equivalent view
  of the GLM, which emphasizes the dependencies between a particular
  time window of stimulus and spike history and conditional intensity
  $\lambda(t)$, which governs the probability of a spike in the
  current time bin (dark gray box).}
\label{fig:modeldiagram}
\end{figure}

The GLM is parametrized by a stimulus filter $\vec k$, which describes
how the neuron integrates an external stimulus, a post-spike filter
$\vec h$, which captures the influence of spike history on the current
probability of spiking, and a scalar $\mu$ that determines the
baseline spike rate.  (See Figure~\ref{fig:modeldiagram}.)  The outputs of
these filters are summed and passed through a nonlinear function $f$
to determine the conditional intensity $\lambda(t)$:
\begin{equation}
\lambda(t) = f(\vec k \cdot \vec x(t)\; +\; \vec h \cdot \vec y_{hist}\left(t\right)\; +\; \mu),
\end{equation}
where $\vec x(t)$ is the (vectorized) spatio-temporal stimulus,
$\vec y_{hist}(t)$ is a vector representing spike history at time $t$, and $f$ is
a nonlinear function that ensures the spike rate is non-negative.
Spikes are generated according to a conditionally Poisson process
\citep{Perkel67a,Cox80}, so spike count $y(t)$ in a time bin of size
$\Delta$ is distributed according to a Poisson distribution:
\begin{equation}
  P(y(t) | \lambda(t))
= \frac{1}{y(t)!} \Big(\Delta \lambda(t)\Big)^{y(t)}
  e^{-\Delta \lambda(t)}.
\end{equation}
In this study we set $f$ to be exponential, although similar
properties can be obtained with other nonlinearities such as the
soft-rectification function.

Unlike classical deterministic models like Hodgkin-Huxley and
integrate-and-fire, the GLM is fundamentally stochastic due to the
assumption of conditionally Poisson spiking. However, this
stochasticity is helpful for fitting purposes because it assigns
graded probabilities to firing events and allows for likelihood-based
methods for parameter fitting \citep{Paninski04NC,Pillow05}.  In fact,
the Poisson GLM comes with a well-known guarantee that the
log-likelihood function is concave for suitable choices of
nonlinearity $f$ \citep{Paninski04}. This means we can be assured of
approaching a global optimum of the likelihood function via gradient
ascent, for any set of stimuli and spike trains (barring any numerical
issues that may complicate achieving the actual maximum for certain datasets, cf.\
\citep{Zhao10}).  This guarantee does not hold for stochastic
formulations of most nonlinear biophysical models, including the
Izhikevich model.  Moreover, despite its stochasticity, the GLM can
produce highly precise and repeatable spike trains in certain
parameter settings, as we will demonstrate below.

\section{GLMs capture a wide array of complex dynamical behaviors}

We fit GLMs to data simulated from Izhikevich neurons set up to
exhibit a range of different qualitative response behaviors. In the
following, we describe these behaviors in detail, beginning with
simpler behaviors, such as tonic spiking and bursting (which have already been demonstrated in previous
work, e.g., \citep{Gerstner92,Jolivet06}) in order to build intuition
for the GLM's basic capabilities, and then move on to more complex
behaviors (such as bistability) and questions of spike timing
reliability and precision.

\subsection{Tonic spiking}
 
\begin{figure}[b!]
\includegraphics[width=.9\textwidth]{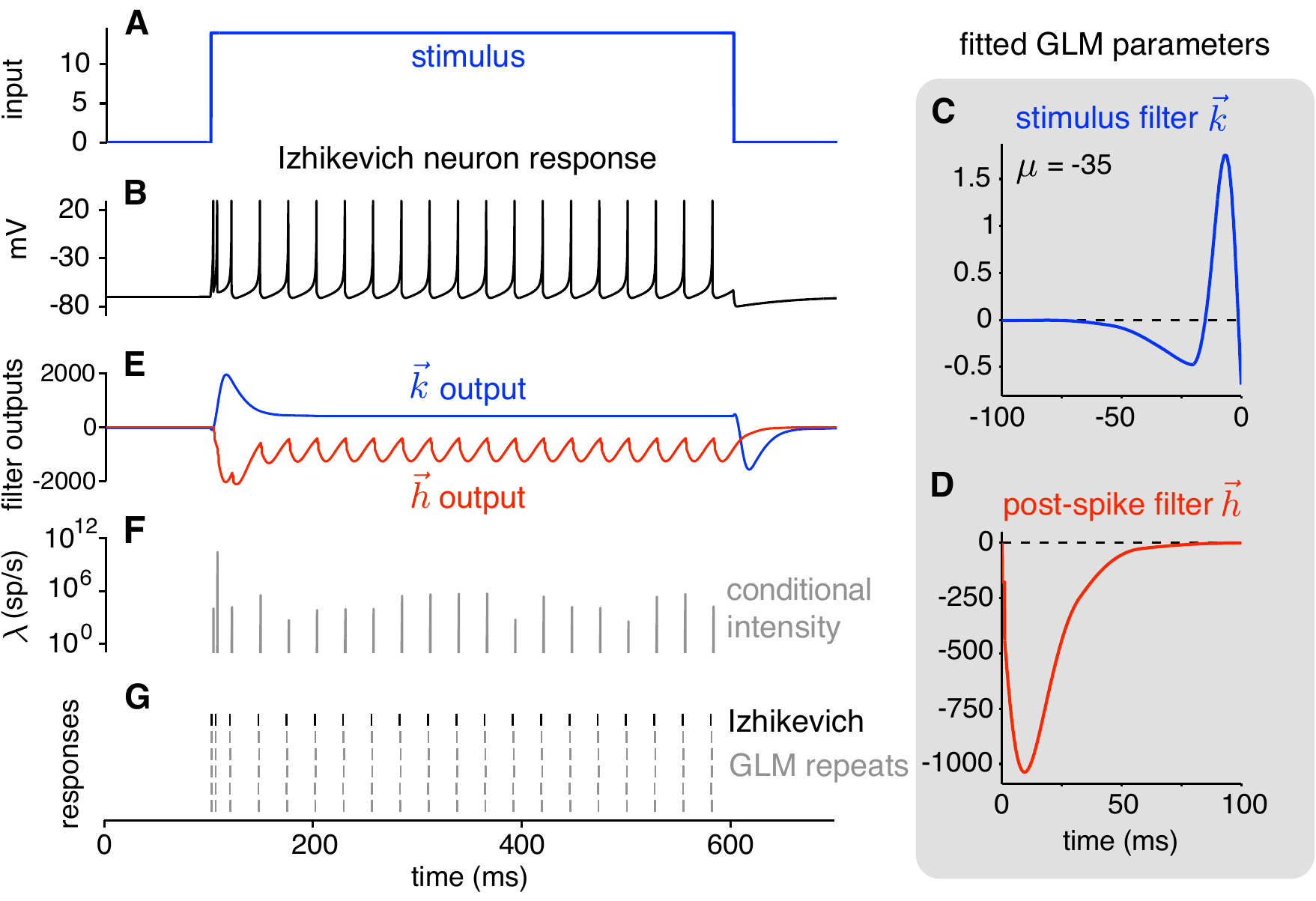}       
\caption{{\bf Tonic spiking behavior.}  A: A step current
  stimulus.  B: Voltage response of the simulated Izhikevich
  neuron.  C: The fitted GLM stimulus filter $\vec k$ has a
  biphasic shape that gives the model a vigorous response to stimulus
  onset and a net positive response to a sustained input.  D:
  The fitted GLM post-spike filter $\vec h$ has a negative lobe that
  imposes strong refractoriness on a timescale of $\approx$50
  ms.  E: Stimulus (blue) and post-spike (red) filter outputs
  during simulated response of the fitted GLM to the stimulus shown
  above on a single trial.  F: The summed filter outputs are
  passed through an exponential nonlinearity to determine the
  conditional intensity $\lambda(t)$, shown here for a single
  trial.  G: Spike train of the Izhikevich neuron (black) and
  simulated repeats of the GLM (gray). GLM spike responses are
  slightly different on each trial, due to the stochasticity of spike
  generation, but reproduce Izhikevich model spike times with high
  precision.}
\label{fig:rs}
\end{figure}

We first examined an Izhikevich neuron tuned to exhibit tonic spiking
(Figure~\ref{fig:rs}A-B; see Table 1 for parameters).  When presented with a
step input current, the Izhikevich neuron responds with a few
high-frequency spikes and then settles into a regular firing pattern
that persists for the duration of the step (Figure~\ref{fig:rs}B). This
response pattern resembles that of a deterministic Hodgkin-Huxley or
integrate-and-fire neuron, albeit with an added transient burst of
spikes at stimulus onset.

We simulated the Izhikevich neuron's response to a series of step
currents and used the resulting training data to perform
maximum-likelihood fitting of the GLM parameters $\{\vec k, \vec h, \mu\}$.
The estimated stimulus filter $\vec k$ is biphasic (Figure~\ref{fig:rs}C),
resulting in a large transient response to stimulus onset, and has a
positive integral, ensuring a sustained positive response to a current
step (Figure~\ref{fig:rs}E, blue trace).  The estimated post-spike filter
$\vec h$ (Figure~\ref{fig:rs}D), by contrast, has a large negative lobe that
provides recurrent inhibition after every spike, enforcing a strong
relative refractory period.  The stimulus filter and post-spike filter
output (shown together for a single trial in Figure~\ref{fig:rs}E) are summed
together and exponentiated to obtain the conditional intensity
$\lambda(t)$ (Figure~\ref{fig:rs}F), also known as the instantaneous spike
rate. 

For this stimulus, the intensity rises very quickly once $\vec h$ decays,
which occurs approximately 50 ms after the previous spike. Note that
the output of the stimulus filter is identical on each trial, whereas
the the output of the post-spike filter varies from trial to trial
because of variability in the exact timing of spikes.  However,
because the rising phase of the conditional intensity is so rapid, spiking
is virtually certain within a small time window sitting at a fixed
latency after the previous spike time.  The combination of strong
excitatory drive from the stimulus filter and strong suppressive drive
from the post-spike filter produces precisely timed spikes across
trials, allowing the GLM to closely match the deterministic firing
pattern of the Izhikevich neuron (Figure~\ref{fig:rs}G).

\subsection{Bursting}

We next examined multi-spike bursting, a more complex temporal
response pattern that requires dependencies beyond the most recent
interspike interval.  
\begin{figure}[h!]
\vspace{10pt}
\includegraphics[width=.85\textwidth]{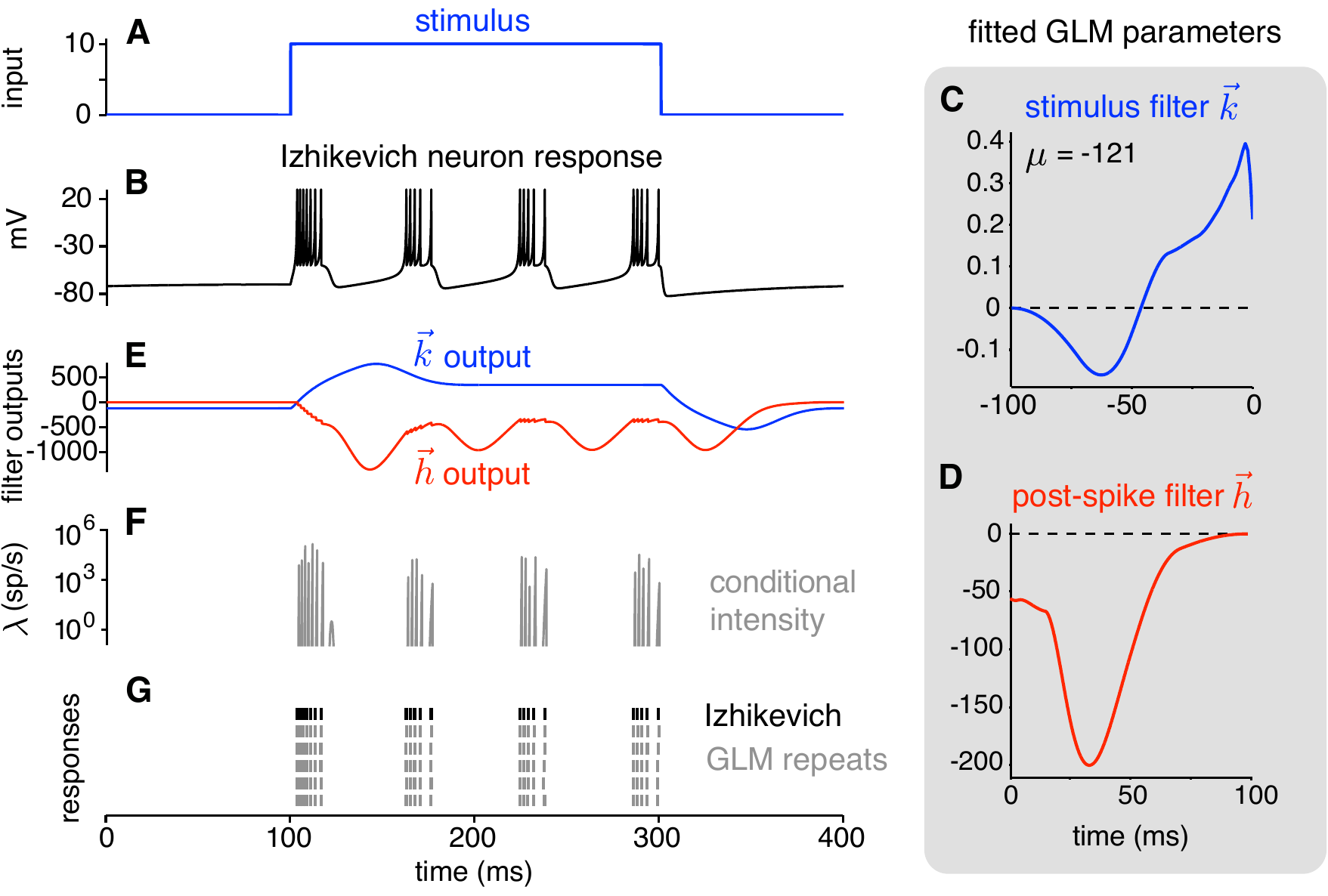} 
\vspace{-5pt}
\caption{{\bf Bursting behavior.} 
A: Step current stimulus.  B: Voltage response of Izhikevich
  neuron.  C: Fitted GLM stimulus filter.  D: Fitted GLM post-spike filter, which creates refractoriness on short
  timescales (within each burst) due to instantaneous depolarization
  following a spike. The large negative lobe $\approx$25-50 ms after a
  spike terminates bursting and strongly suppresses firing between
  bursts.  E: Stimulus (blue) and post-spike (red) filter
  outputs for simulated response of the GLM to step current shown
  above on a single trial.  F: Output of the nonlinearity (conditional
  intensity) $\lambda(t)$ on a single trial.  G: Spike train
  of Izhikevich neuron (black) and simulated repeats of the fitted GLM
  (gray).}
\label{fig:tb}
\end{figure}
Once again, we simulated responses from an
Izhikevich neuron tuned to exhibit tonic bursting (Figure~\ref{fig:tb}A-B)
and used the resulting data to fit GLM parameters (Figure~\ref{fig:tb}C-D).
The estimated stimulus filter $\vec k$ is biphasic with a larger positive
than negative lobe, which drives rapid spiking at stimulus onset and
generates sustained drive during an elevated stimulus (Figure~\ref{fig:tb}E,
blue trace).  The post-spike filter $\vec h$ has an immediate negative
component that creates a relative refractory period after each spike,
and an even more negative mode after a latency of $\approx$40 ms; the
accumulation of these negative components over multiple spikes gives
rise to a sustained suppression of activity between bursts.

The GLM captures the bursting behavior of the Izhikevich neuron with
high precision, including the fact that the first burst after stimulus
onset contains a different pattern of spikes than subsequent bursts.
This difference arises from precise interactions between the stimulus
and post-spike filter outputs.  During the first burst, fast spiking
arises from an interplay between monotonically increasing stimulus
filter output (Figure~\ref{fig:tb}E, blue) and tonic decrements induced by the
post-spike filter after each spike (Figure~\ref{fig:tb}E, red).  After each
spike, the post-spike filter reduces the conditional intensity by a
fixed decrement, but this decrement is soon overwhelmed by the rising
wave of input from the stimulus filter, which creates a rapid rise and
fall of the conditional intensity time-locked to each Izhikevich
neuron spike time.  The pattern continues until accumulated
contributions from the delayed negative lobes of post-spike filter
overwhelm those from the stimulus filter and the burst terminates.
Subsequent bursts are governed by a somewhat different interplay
between stimulus and post-spike filter outputs: bursts sit on a rising
phase of the conditional intensity due to the removal of suppression
from the previous burst. This rise is more gradual than the drive
induced by stimulus onset, and results in bursts with longer
inter-spike intervals and fewer spikes per burst, but the resulting
spike pattern is nonetheless captured with high precision and
reliability from trial to trial.

\subsection{Bistability}

\begin{figure}[b!]
\includegraphics[width=.85\textwidth]{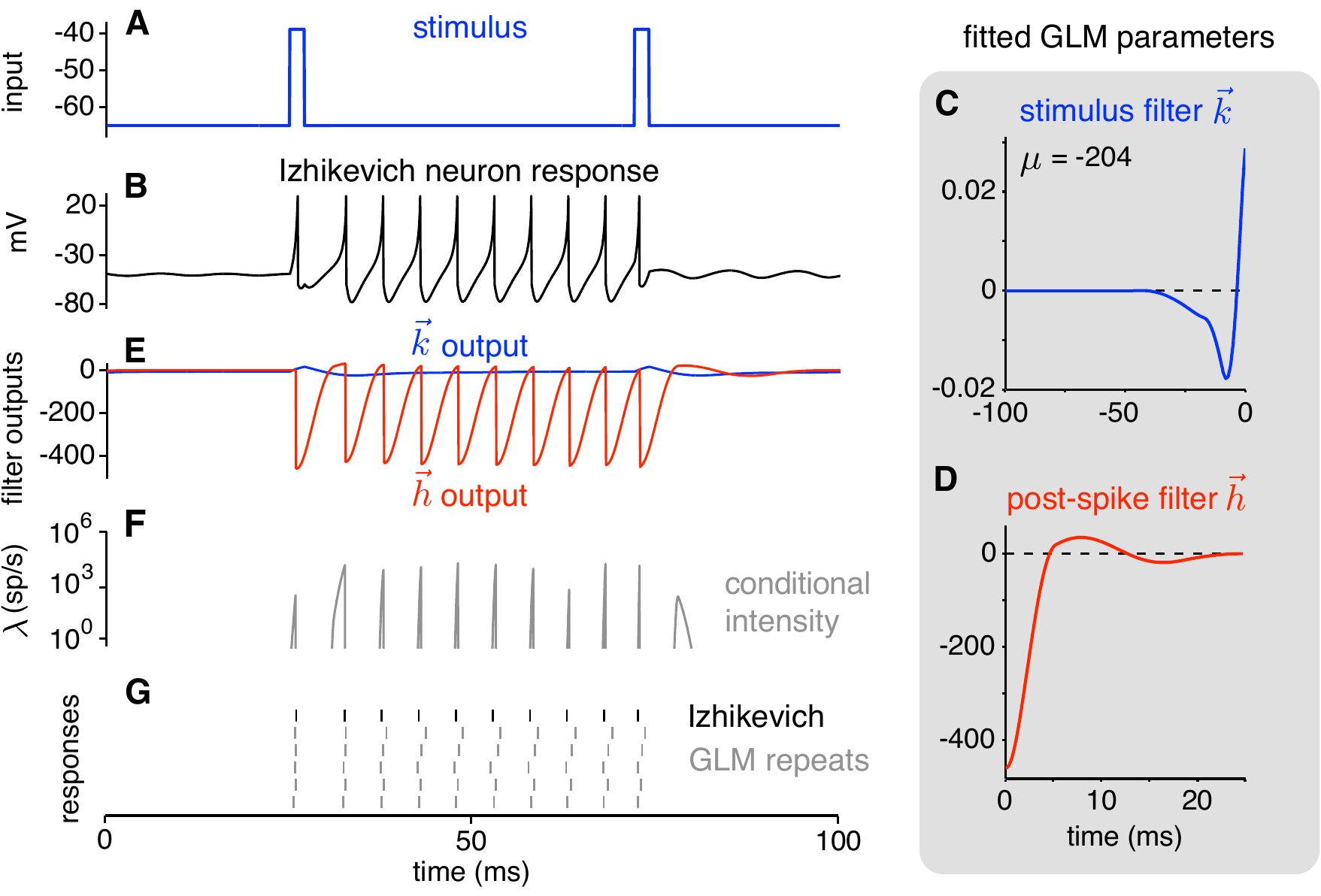} 
\caption{{\bf Bistable responses.} 
A: Stimulus consisting
  of two brief positive current pulses.  B: Voltage response
  of Izhikevich neuron, which exhibits bistability. The first pulse
  initiates a tonic spiking mode and the second pulse (precisely timed
  to the phase of the spike response) terminates it, returning to a
  quiescent mode. C: Fitted GLM stimulus filter, which
  provides a biphasic impulse response.  D: Fitted
  post-spike filter, which imposes a refractory period of $\approx$4
  ms and gives increased probability of firing $\approx$5-10 ms after
  each spike.  E: GLM stimulus (blue) and post-spike (red)
  filter outputs on a single trial. F: Output of the
  nonlinearity gives the conditional intensity for a single
  trial.  G: Spike train of Izhikevich neuron (black) and
  simulated repeats of the fitted GLM (gray).}
\label{fig:bistable}
\end{figure}

Bistability refers to the phenomenon in which there are multiple
stable response modes for a single input condition. A common form of
bistability observed in real neurons is the ability to inhabit either
a tonically active state or a silent state for a given level of
current injection.  The Izhikevich model can exhibit this form of
bistability, wherein a brief positive current pulse is sufficient to
kick it between states: the neuron can inhabit a silent state in the
absence of stimulation, but a brief positive current pulse kicks it
into a tonically active state, and an appropriately-timed positive
pulse kicks it back to the silent state (Figure~\ref{fig:bistable}A-B).

We fit a GLM to spike trains simulated from such a bistable Izhikevich
neuron and found that the fitted GLM can capture bistable behavior of
the original model with high accuracy (Figure~\ref{fig:bistable}C-G).  When
stimulated during the silent state, the GLM emits a spike due to the
positive output of the stimulus filter (Figure~\ref{fig:bistable}E, blue), and
tonic firing ensues due to a positive lobe in the post-spike filter
that causes self-excitation at a fixed latency of approximately 10
ms after the previous spike (Figure~\ref{fig:bistable}E).  The GLM returns
from the active state to the silent state when a positive stimulus
pulse synchronizes negative lobes of the stimulus and post-spike
filters. Only the combination of these two negative drives is strong
enough to shut off spiking; without the negative drive created by
previous spikes (appearing in the post-spike filter at a latency of
15 ms after a spike), suppression from the negative lobe of the
stimulus filter is not strong enough to prevent spiking.  As with the
tonic spiking neuron discussed above, the interaction between the
stimulus and post-spike filters generates rapid rises in the
conditional intensity (Figure~\ref{fig:bistable}F), leading to precisely timed
spikes that mimic those of the deterministic Izhikevich neuron
(Figure~\ref{fig:bistable}G).

This Izhikevich neuron (with the same parameters) can also exhibit a second form of bistability,
in which the return to the silent state from the active state is
induced by a negative instead of a positive current pulse. We
performed a similar fitting exercise and found that the GLM is also
able to reproduce this behavior.  Firing
is initiated and maintained by a similar mechanism as the first form
of bistability, but firing offset occurs due to the fact that a
negative stimulus pulse creates immediate negative output from the
stimulus filter, which suppresses firing during the time when a spike
would have occurred due to spike-history filter input.  Tonic firing
is extinguished more rapidly in this second form of bistability than
the first (see Figure~\ref{fig:allbehaviors} below).

\subsection{Type I and type II firing}

\begin{figure}[b!]
\includegraphics[width=.8\textwidth]{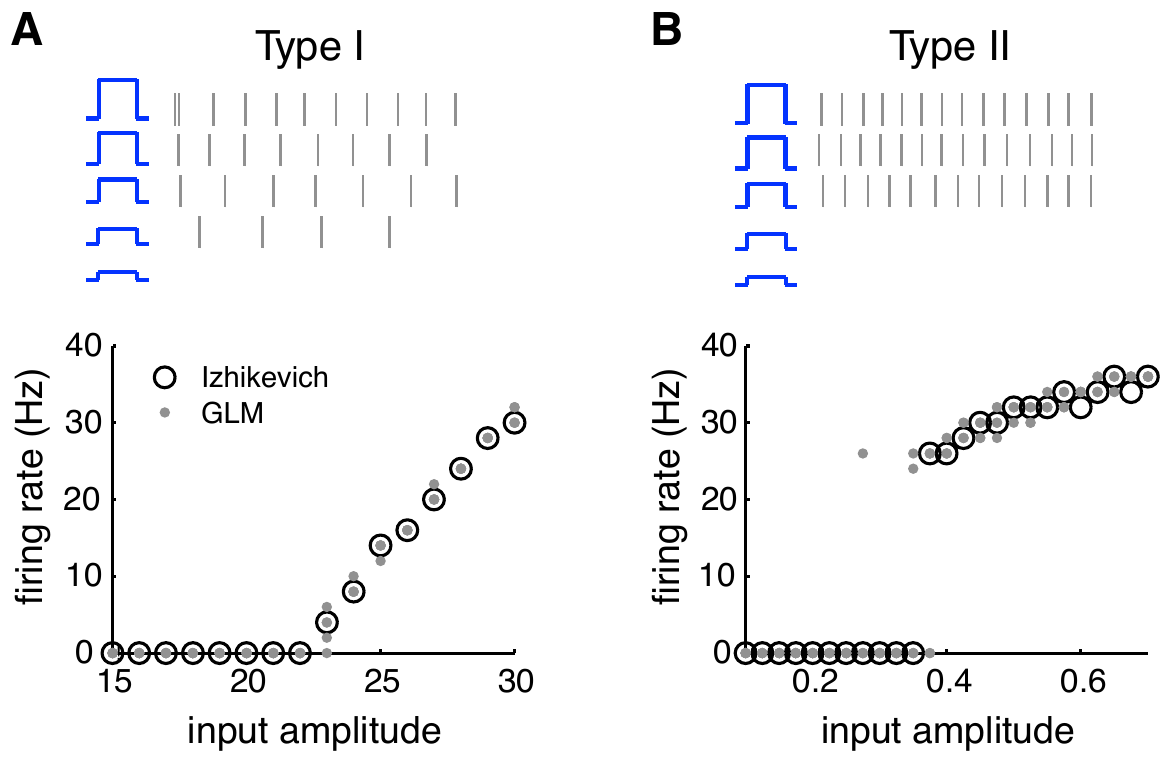}  
\caption{{\bf Type I and type II firing curves.}  
A: {\it{Top:}} Example responses of a GLM exhibiting type I firing
  behavior.  The spike rate increases continuously from zero in
  response to current steps of increasing amplitude.  {\it{Bottom:}}
  F-I curve for a type I Izhikevich neuron (black) and corresponding
  GLM (gray). For the GLM, responses are plotted for five repetitions
  of each input amplitude.  B: Similar plots for type II
  firing behavior, characterized by a discontinuous jump from zero to
  a finite spike rate in responses to current steps of increasing
  amplitude.}
\label{fig:class1class2}
\end{figure}

Neurons have been classified as exhibiting either type I or type II
dynamics based on the shape of their firing rate vs. intensity (F-I)
curve.  Type I neurons can fire at arbitrarily low rates for low
levels of injected current, whereas type II neurons have discontinuous
F-I curves that arise from an abrupt transition from silence to a
finite non-zero firing rate as the level of injected current increases
\citep{Hodgkin1948}.  We simulated Izhikevich neurons that exhibit each
of these response types using published parameters.  Inputs consisted
of 500 ms current steps of varying amplitude.  The resulting F-I
curves for the Izhikevich type I and type II neurons are shown in
black in Figure~\ref{fig:class1class2}A and Figure~\ref{fig:class1class2}B,
respectively.  We fit GLMs using data from each Izhikevich neuron and
found that the fitted GLMs capture the two response types with high
temporal precision.  The corresponding F-I curves are shown in gray in
Figure~\ref{fig:class1class2} and accurately mimic the behaviors of the
Izhikevich neuron.  Similar F-I curves have been demonstrated previously in \citep{Gerstner92} and \citep{Mensi12}.

The only discrepancy between the Izhikevich and GLM neurons occurs for
the type II cell at input amplitudes near the Izhikevich neuron's
threshold.  On some trials when the input amplitude falls below this
threshold, the GLM jumps into a a tonic firing state for the duration
of the stimulus.  Similarly, on some trials when the input amplitude
falls above this threshold, the GLM fails to initiate firing.  This is
unsurprising given the stochastic nature of the GLM.  Importantly, the
GLM never fires at a low rate, but rather abruptly transitions from no
firing to firing at a baseline level of $\approx$25 Hz, reflecting type
II behavior.

\subsection{Additional behaviors}

\begin{figure}[h!]
\includegraphics[width=1\textwidth]{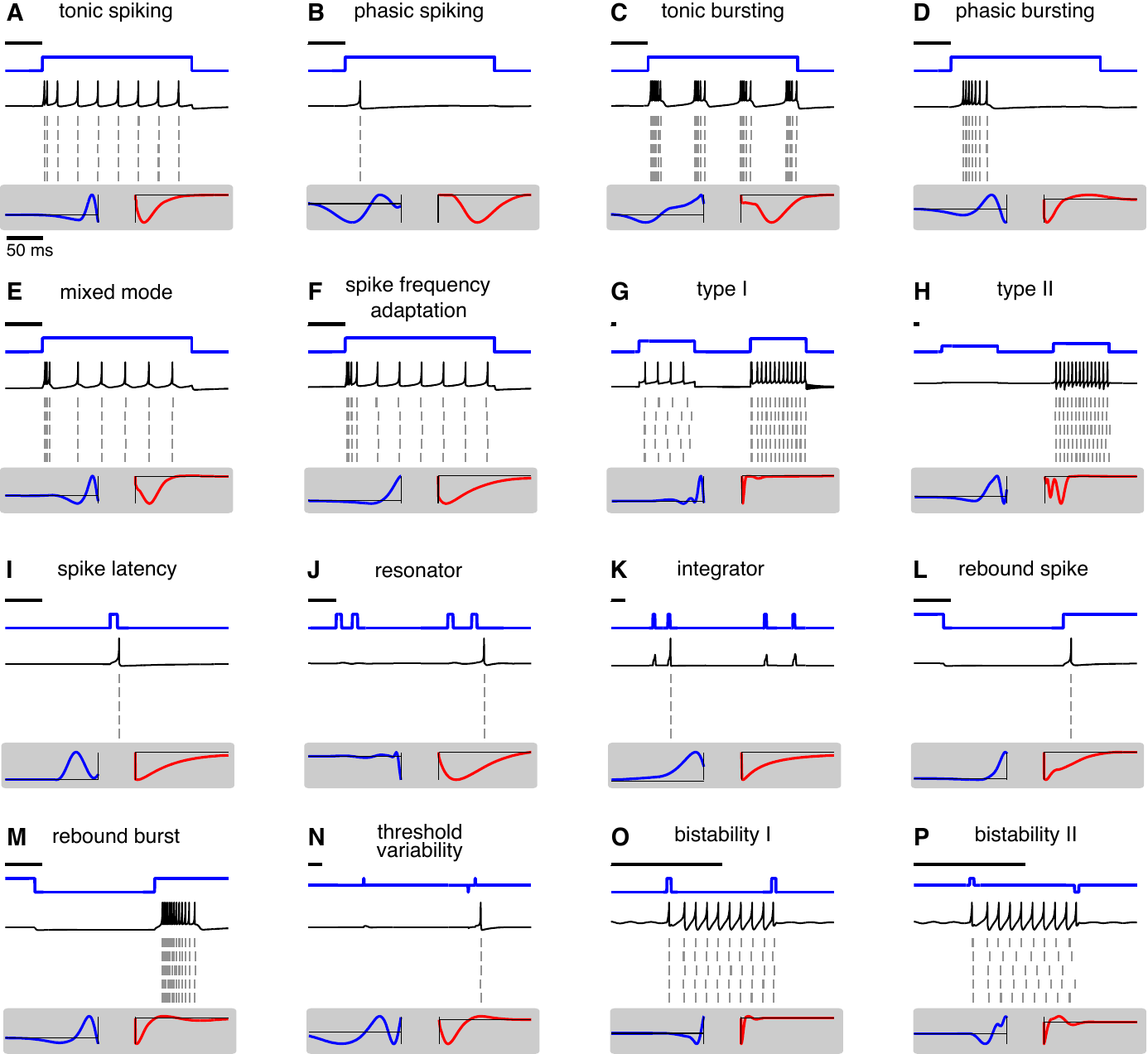} 
\caption{{\bf Suite of dynamical behaviors of Izhikevich and GLM
    neurons.} 
    Each panel, top to bottom: stimulus (blue), Izhikevich
  neuron response (black), GLM responses on five trials (gray),
  stimulus filter (left, blue), and post-spike filter (right,
  red). Black line in each plot indicates a 50 ms scale bar for the
  stimulus and spike response. (Differing timescales reflect
  timescales used for each behavior in original Izhikevich paper
  \citep{Izhikevich04}). Stimulus filter and post-spike filter plots
  all have 100 ms duration.}
\label{fig:allbehaviors}
\end{figure}

We fit GLMs to every dynamical behavior considered in
\citep{Izhikevich04} with the exception of purely subthreshold
behaviors, since GLM fitting uses spike trains and does not consider
sub-threshold responses.  The full suite of behaviors is shown in
Figure~\ref{fig:allbehaviors}, with responses of the Izhikevich neurons in
black and spike responses of the GLM in gray.  This list includes tonic and phasic spiking, tonic and phasic
bursting, mixed mode firing, spike frequency adaptation, type I and
type II excitability, two different forms of bistability, and several
others that depend primarily on the shape of stimulus filter. Several
additional behaviors that can be captured by a GLM are not depicted
in Figure~\ref{fig:allbehaviors} as they can be achieved by a trivial
manipulation of the stimulus filter; for example, inhibition-induced
bursting can be achieved by simply flipping the sign of the stimulus
filter for the bursting neuron shown in Figure~\ref{fig:allbehaviors}C.
Previous work has shown the Izhikevich neuron to be capable of
producing 18 distinct spiking behaviors \citep{Izhikevich04}, and we
found that all can also be produced by a GLM.

\subsection{Systematic variation of filter amplitudes}

\begin{figure}[h!] 
\includegraphics[width=1\textwidth]{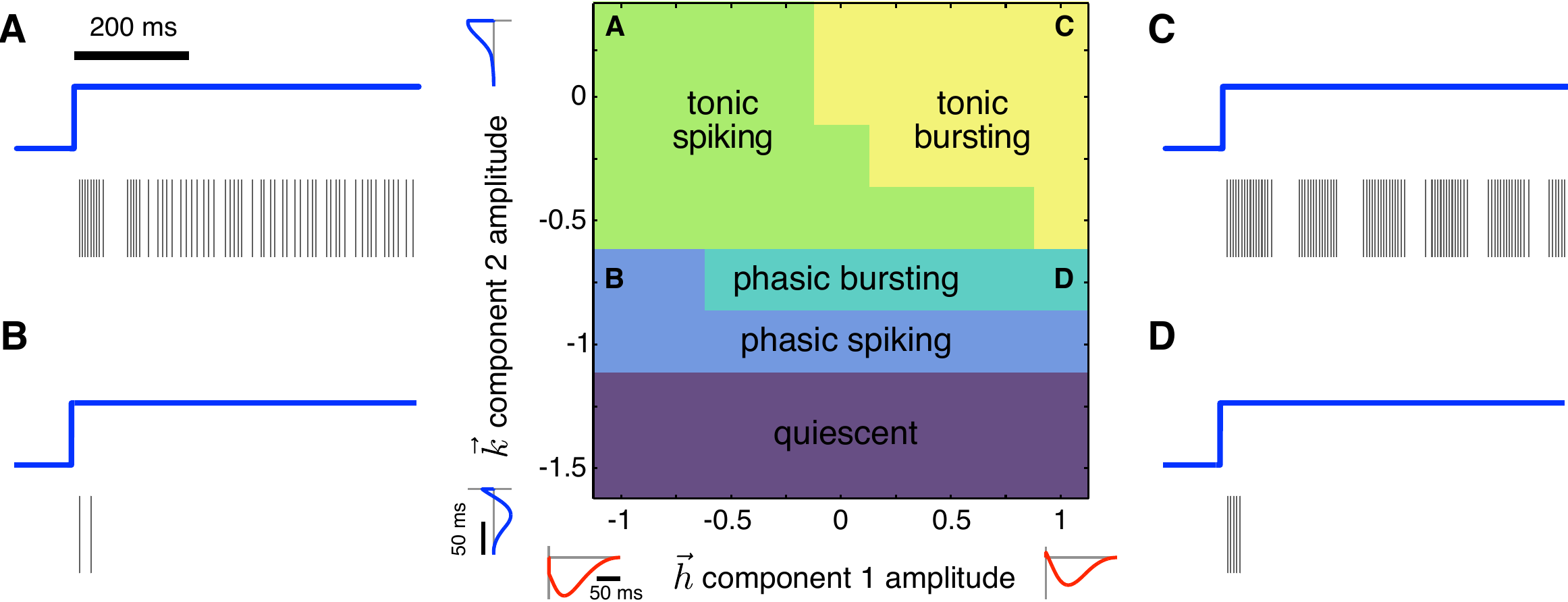} 
\caption{{\bf Changes in a single component of each filter can produce a variety of behaviors.} Center: Amplitudes for a single component of the stimulus filter (ordinate) or post-spike filter (abscissa) were varied. Responses were simulated for 15 trials of a step stimulus, and the most common behavior produced is indicated by color. Small panels show example filters at each extreme of the range tested. A-D: Example responses (gray) to step stimulus (blue) for GLMs with filters indicated by corresponding letter in center panel.}
\label{fig:param_sweep}
\end{figure}

We next considered what happens to the behaviors produced by a GLM as
some aspect of the filters is systematically varied.  To do so, we
created stimulus and post-spike filters composed by linear
combinations of two basis filters, and then systematically varied the
amplitude of one basis filter while holding the other fixed.  (See
Methods for details.)  Figure~\ref{fig:param_sweep} shows the phase
space of qualitative spiking behaviors obtained at different points in
this 2D filter space.

When the stimulus filter has a strongly negative component (center panel, bottom), a positive stimulus pulse does not produce enough driving force to cause the neuron to spike at all (quiescent).  As the amplitude of this component of the stimulus filter is increased, the neuron receives stronger and stronger input and is driven first to spike once or twice (phasic spiking), and eventually to emit a burst of spikes (phasic bursting).  The stimulus filter largely drives changes between these behaviors, with the additional detail that a strongly negative post-spike filter component is able to inhibit a burst that would otherwise occur (upper left corner of ``phasic spiking" region).  As the stimulus filter component becomes still more positive and the stimulus filter transitions from being biphasic to more monophasic, it produces a positive driving force for the duration of the stimulus step, rather than just the onset.  This causes the neuron to fire for the duration of the step.  

Importantly, the post-spike filter here determines the nature of this sustained firing.  A post-spike filter that is purely negative beginning at short timescales (top, left side) mimics a relative refractory period, inhibiting additional spikes for a short window following each elicited spike and resulting in tonic spiking.  If, on the other hand, the post-spike filter is only weakly negative at short timescales while being more strongly negative at longer timescales, this creates multiple timescales in the neuron's response (top, right side).  At short timescales, there is little inhibition from each spike (beyond the absolute refractory period), so additional spikes may occur.  Over longer timescales, inhibition is accumulated over multiple spikes, which eventually shuts off spiking.  After a brief window of no spikes, the inhibition is relaxed and spiking commences again until enough inhibition is accumulated to shut spiking off.  This cycle results in tonic bursting for the duration of the step.  In the extreme case where the post-spike filter is actually biphasic (top, far right), each spike promotes additional spikes on short timescales, leading to highly regular timing of spikes within bursts (Figure~\ref{fig:param_sweep}C).  This set of behaviors could be achieved by simply sweeping over the amplitude of a single basis vector in each filter.  Incorporating shifts to the basis vectors or additional basis vectors would likely be necessary to achieve more complex behaviors, such as bistability.

Although we have drawn clear borders at the transition between behaviors, these transitions in fact occur gradually.  Near the border between phasic spiking and phasic bursting, for example, there will be some trials where a single spike is produced and other trials where a burst is elicited.  We have indicated the behavior that is produced most frequently here for simplicity.  The transition from tonic bursting to tonic spiking also occurs gradually, with the near perfectly regular bursting breaking down into more irregular firing until no apparent bursts are produced.  If the post-spike filter is made even more negative than the range explored in this figure, the timing of tonic spiking becomes near perfectly regular as well.  This is easily explained by the fact that as the post-spike filter becomes more and more negative, it imposes stronger refractoriness on the cell, which results in more regular spike timing.  As the post-spike filter component amplitude is changed, there is therefore a gradual change from precisely timed bursts, to irregular firing, to precisely timed tonic spiking.  In the following section, we further explore questions of spike timing precision in the GLM.

 \subsection{Generalization to new stimuli}

\begin{figure}
\hspace{-70pt}
\floatbox[{\capbeside\thisfloatsetup{capbesideposition={right,top},capbesidewidth=5cm}}]{figure}[\FBwidth]
{\caption{{\bf GLMs have limited ability to genearlize to new stimuli.} A: Inputs used to generate responses from Izhikevich neurons. Step amplitudes were 7, 14, \& 28; 0.3, 0.6, \& 1.2; and 5, 10, \& 20 for B, C, \& D, respectively.  Standard deviation of noise was 7.2 for all neuron types.  B: {\it Top:} Responses of a regular spiking Izhikevich neuron to the above stimuli.  {\it Middle:} Spike responses of the Izhikevich neuron (red) and GLM (black) fit on responses to only the middle step size (indicated by gray box).  {\it Bottom:} Spike responses of the Izhikevich neuron (red) and GLM (black) fit on responses to all three step sizes (indicated by gray box). C: Responses of a phasic bursting Izhikevich neuron. All panels as in B.  D: Responses of a tonic bursting Izhikevich neuron.  All panels as in B \& C. }
\label{fig:gen}}
{\includegraphics[width=.75\textwidth]{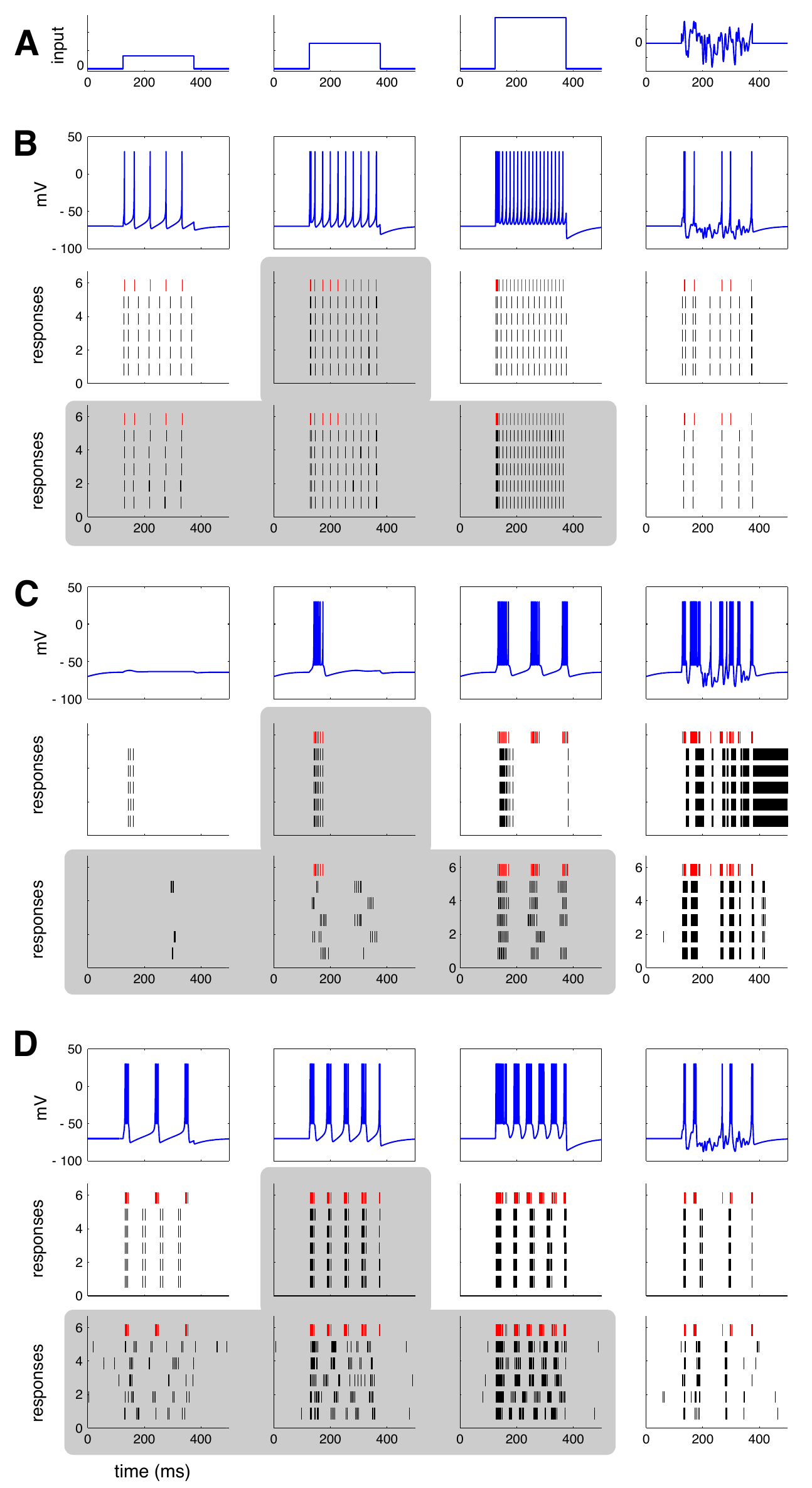}} 
\end{figure}

The behaviors shows in Figure~\ref{fig:allbehaviors} were all fit
using stimuli that probe only a small range of the possible behaviors
of the neuron.  For example, many were probed using only a single step
height.  A natural question that arises is therefore: how well will
these fitted GLMs generalize to predict the Izhikevich neuron
responses to new stimuli?  We examined this question for three
canonical  Izhikevich neurons from our study
(Figure~\ref{fig:gen}).  We first generated responses from a regular
spiking Izhikevich neuron using three step heights and one noise
stimulus (Figure~\ref{fig:gen}A; B, \emph{top}).  We then simulated
responses to these stimuli using the GLM fit only to the intermediate
step height (Figure~\ref{fig:gen}B, \emph{middle}).  (This is the same
fit as Figure~\ref{fig:rs}.)  The GLM responses nearly perfectly
capture the Izhikevich responses for the original stimulus, and the
GLM maintains regular firing patterns for the other step heights.
However, the GLM's firing rate is too high for the small step and too
low for the large step.  While the GLM accurately captures some firing
events for the noise stimulus, the firing rate is overall too high.
We next refit a GLM to responses from all three step heights, using
the same set of basis vectors as the original fit
(Figure~\ref{fig:gen}, \emph{bottom}).  The responses to all three
step heights are captured nearly perfectly.  Additionally, the
response to the noise stimulus is much more similar to that of the
Izhikevich neuron.  Although there is one firing event that occurs at
a delay, the regular spiking GLM fit on an enriched stimulus set is
better able to generalize to the noise stimulus.

We performed this same test for neurons showing phasic bursting (Figure~\ref{fig:gen}C) and tonic bursting (Figure~\ref{fig:gen}D).  (Note that although some responses of the phasic bursting Izhikevich neuron are not actually phasic, we retain the naming convention given to this set of parameters in the original paper.)  For phasic bursting, the GLM fit on additional step sizes improves the accuracy of responses to the smallest and largest steps, while decreasing the accuracy of responses for the original step of intermediate size.  There is marked improvement in the accuracy of responses to the noise stimulus, with many firing events being accurately captured and the GLM no longer exhibiting runaway excitation.  For tonic bursting, the refit GLM retains bursting behavior but fails to even capture responses for the steps on which it was trained.  As noted above, when refitting we used the same set of basis vectors as the initial fits for fair comparison.  It is possible that by increasing the number of basis vectors used or tuning their properties that better fits to all stimuli might be achieved.

Taken together, these results show that while GLMs might retain some characteristic response features (such as bursting) when probed with new stimuli, they often have limited ability to generalize beyond stimuli on which they are directly fit.


\section{Spiking precision and reliability}

A noteworthy feature of the spike trains of the GLM neurons considered
above is their high degree of spike timing precision and reliability
across trials.  This precision arises from the fact that the
conditional intensity (or instantaneous spike rate) rises abruptly at
spike times (due to filter outputs passing through a rapidly
accelerating exponential nonlinearity), and decreases immediately
after each spike due to suppressive effects of the post-spike
filter. By contrast, a Poisson GLM without recurrent feedback, more
commonly known as a linear-nonlinear Poisson (LNP) cascade model,
cannot produce temporally precise spike responses to a constant
stimulus because its output is constrained to be a Poisson process.

\begin{figure}[h!]
\vspace{10pt}
\includegraphics[width=.7\textwidth]{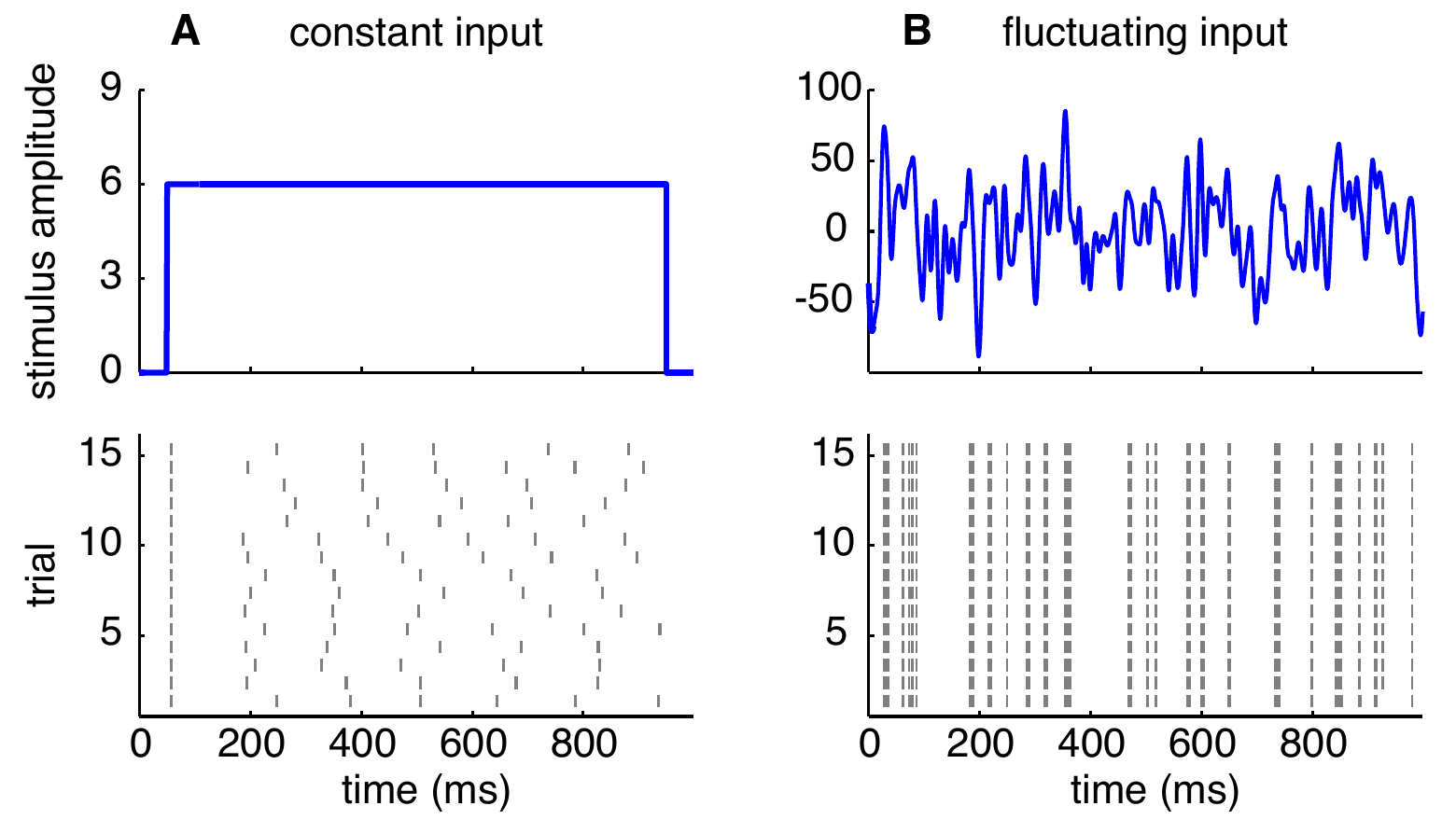} 
\vspace{-5pt}
\caption{{\bf Stimulus-dependent spike timing reliability.}
A: {\it Top:} Weak step stimulus. {\it Bottom:} Spike
  train responses of tonic-spiking GLM on 15 repeated trials. Although
  the first spike is precise and reliable, subsequent spikes have
  irregular timing from trial to trial.  B: {\it Top:}
  Rapidly fluctuating stimulus. {\it Bottom:} Spike response of same
  GLM neuron on 15 repeated trials, exhibiting a high degree of
  precision and reliability. (Compare to Figure 1 of \citep{Mainen95}.)}
\label{fig:mainen}
\end{figure}

Real neurons, however, seem to be capable of both response modes: they
emit precisely-timed spikes in some settings and highly variable spike
trains in others.  A seminal paper by Mainen \& Sejnowski illustrated
this duality by showing that spike responses to a constant DC current
exhibit substantial trial-to-trial variability, whereas responses to a
rapidly fluctuating injected current are precise and repeatable across
trials \citep{Mainen95}. Deterministic models like the Izhikevich model
cannot, of course, mimic this property because their spikes are
perfectly reproducible for any stimulus.  (A stochastic version of the
Izhikevich model with an appropriate level of injected noise could
likely overcome this shortcoming, however.  See \cite{Schneidman98} for a similar case in a Hodgkin-Huxley neuron.)  Here we show that the GLM
naturally reproduces the same form of stimulus-dependent changes in
precision and reliability observed in real neurons.  Figure~\ref{fig:mainen}
shows that a single GLM (with parameters identical to those fit to the
tonic-spiking Izhikevich neuron, shown in Figure~\ref{fig:rs}) produces
irregular spiking in response to a constant stimulus with
low-to-intermediate amplitude, and precisely-timed, reliable spikes in
response to a stimulus with large, rapid fluctuations.

\section{GLMs can produce super-Poisson variability}

We have shown that GLM neurons can reproduce the high degree of spike
timing precision found in real neurons stimulated with injected
currents.  However, a variety of studies have reported that neurons
exhibit {\it overdispersed} responses, or greater-than-Poisson spike
count variability in response to repeated presentations of a sensory
stimulus \citep{Tomko74,Tolhurst83,Shadlen98,Goris14}.  A prominent
recent study from Goris, Movshon, \& Simoncelli showed that the degree
of overdispersion grows with mean spike count, so that the Fano factor
(variance-to-mean ratio) is an increasing function of spike rate
\citep{Goris14}.  They proposed a doubly stochastic model to account
for this phenomenon, in which the rate of a Poisson process is
modulated by a slowly fluctuating stochastic gain variable $g$. For
each trial, $g$ is drawn from a gamma distribution with mean 1 and
variance $\sigma_g^2$.  (See Methods for details.)

\begin{figure}[b!]
\vspace{10pt}
\includegraphics[width=1.05\textwidth]{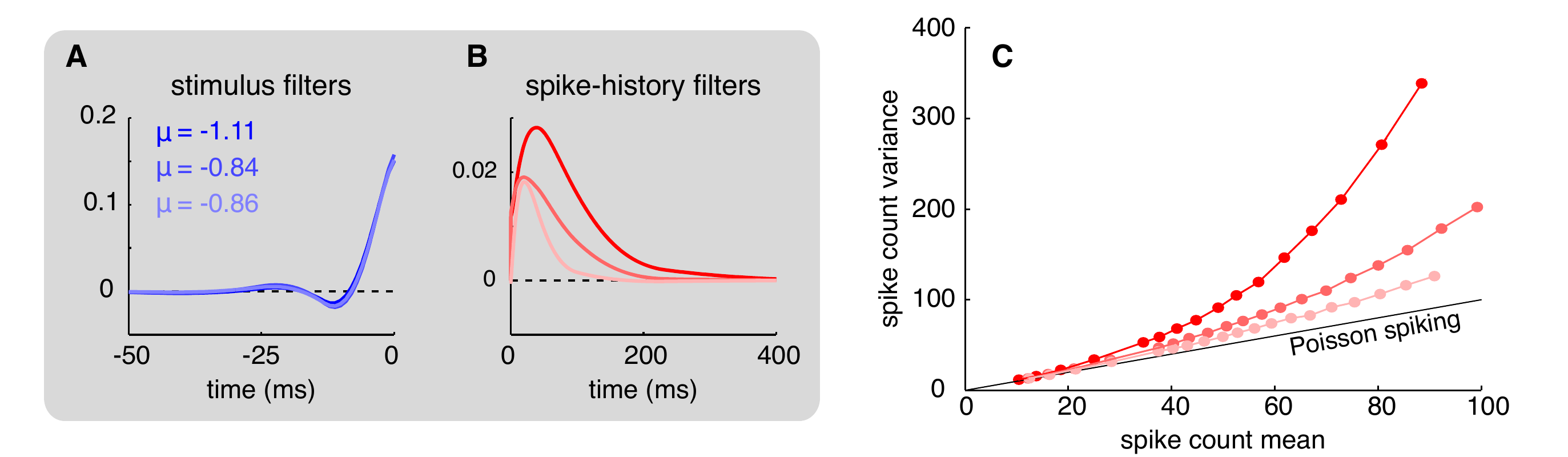}
\caption{{\bf GLMs can produce super-Poisson variability.}
 A: Stimulus filters for GLMs trained on three different
  levels of variability: high (dark blue), medium (medium), and low
  (light) super-Poisson variability. Spike count mean was identical in
  the three cases: 100 spikes/s. B: Post-spike filters for
  high (dark red), medium (medium), and low (light) super-Poisson
  variability. C: Spike count variance versus mean for three
  levels of variability. This relationship is strikingly similar to
  that observed in many cortical neurons.}
\label{fig:goris_fig}
\end{figure}

We sought to determine if a GLM with spike-history dependence can also account for the mean-dependent
overdispersion found in neural responses.  To test this possibility,
we simulated spike trains from the doubly stochastic model of Goris
{\it et al} for three different
settings of the over-dispersion factor $\sigma_g^2$ with the same mean spike rate (100 spikes/s), and fit a GLM to
the spike trains associated with each value of $\sigma_g^2$.  We then simulated responses from each GLM to 500 ms pulses at a number of different input intensities, with each point in Figure~\ref{fig:goris_fig} corresponding to a different intensity.

We found that the GLM can indeed match the qualitative behavior of the
Goris {\it et al} model, giving approximately Poisson responses at low
spike rates and increasingly overdispersed responses at higher rates
(Figure~\ref{fig:goris_fig}C).  To match the data from larger values of
overdispersion factor $\sigma_g^2$ (darker curves in
Figure~\ref{fig:goris_fig}C), the GLM relies on increasing amounts of
self-excitation from the post-spike filter, but exhibits no changes in stimulus
filter (Figure~\ref{fig:goris_fig}A-B).  The filters here do not include an absolute refractory period, as the original model does not incorporate one.  However, similar results can be achieved when a refractory period is enforced in the training data.  This will result in post-spike filters with strongly negative lobes on short timescales, which impose refractoriness, but otherwise similar filters to those in Figure~\ref{fig:goris_fig}.  Unlike models with purely suppressive spike history effects, which capture effects due to refractoriness and reduce variability in firing (e.g., \citep{Berry98,Uzzell04}), here we show that allowing spike history effects to be excitatory can result in increased variability.  While the former might be suitable for early sensory areas, such as the retina, the latter better captures the super-Poisson variability observed in higher visual areas.

Intuitively, the GLM generates overdispersed spike counts because of
dependencies the spike-history filter induces between early and late
spikes during a trial: if the GLM neuron generates a
larger-than-average number of spikes early in a trial, the positive
post-spike filter produces a higher conditional intensity (and hence
more spiking) later in the trial; conversely, if a neuron emits
fewer-than-average spikes early in a trial, the conditional intensity
will be lower later in the trial (yielding less spiking).  The Goris
{\it et al} model can be seen to capture similar dependencies between
early and late spikes via the stochastic gain variable $g$, which is
constant during a trial but independent across trials.  Thus, it is
reasonable to view $g$ in the Goris {\it et al } model as a proxy for
the accumulated self-excitation from spike-history filter outputs
under a GLM.  We note, however, that attempts to drive the GLM to
higher levels of overdispersion (e.g., Fano factors significantly
$> 3$) often resulted in runaway self-excitation, indicating that GLMs
may require additional mechanisms to maintain stability in order to
produce highly over-dispersed responses through recurrent excitation
alone \citep{Gerhard17,Hocker17}.  An alternative mechanism for
generating over-dispersed responses with GLMs is through the addition of
latent stochastic inputs \citep{Rabinowitz15}, an avenue we have not explored here.

\section{Discussion}

We have shown that recurrent Poisson GLMs can capture an extensive set
of behaviors exhibited by biological neurons, including tonic and
phasic spiking, bursting, spike frequency adaptation, type I and type
II behavior, and bistability.  GLMs can also reproduce widely varying
levels of response stochasticity, ranging from precisely timed spikes
with negligible trial-to-trial variability, to substantially
super-Poisson spike count variability.  We have also shown that, like
real neurons, GLMs can exhibit irregular firing in response to a
constant stimulus, but precise and repeatable firing patterns in
response to a temporally varying stimulus.  Thus, generalized linear
models are able to capture a rich array of spiking behaviors like many
dynamical models, while remaining tractable to fit to neural data.

\subsection{Relationship to previous work}
As mentioned above, GLMs have strong connections to a number of other models. It is particularly worth noting the connection between GLMs and generalized integrate-and-fire models, such as the spike response model (SRM) extensively studied by Gerstner and colleagues \citep{Gerstner92,Gerstner01}.  These models draw on much earlier work which incorporated a variable threshold that depends on spiking history \citep{Weiss66,Geisler66}.  The SRM includes a membrane filter (analogous to the stimulus filter here) and both a spike afterpotential and moving threshold (which can be combined and are analogous to the spike history filter here). In its simplest formulation, the SRM is a deterministic model. Although the threshold for spiking can shift as a function of spike history, a spike will occur precisely at each threshold crossing. 

Extensions of the model have incorporated so-called ``escape noise," where spiking no longer occurs deterministically at threshold crossings, but rather the probability of spiking depends on the distance of the membrane voltage to threshold \citep{Gerstner92,Jolivet06}. This variant of the SRM is in fact a GLM, and it is therefore worthwhile to consider how previous work investigating the SRM with escape noise relates to our results here. Early work demonstrated that such a model was capable of producing responses with high temporal precision, including both tonic spiking as well as tonic bursting \citep{Gerstner92}, though demonstrating the range of behaviors that could be produced by the SRM was not the focus of this study. Additional work demonstrated that the model could produce highly repeatable spike trains to a noisy stimulus (similar to Figure~\ref{fig:mainen}B, though no comparison of irregular firing in response to a constant stimulus was shown) \citep{Jolivet06}.  Other studies have shown that the SRM can capture the detailed statistics of neural responses \citep{Mensi12,Pozzorini13}.  Further, many of these studies show that the spike responses model can be used to capture not only the relationship between an external stimulus and a neuron's response, but also to faithfully capture the relationship between intracellularly recorded neural responses and injected current \citep{Jolivet06,Pozzorini13}.

\subsection{Limitations}
Despite their many advantages, GLMs have several limitations that bear further discussion.  First, GLMs
often do not generalize well across stimulus distributions; models fit
with a particular set of stimuli often do not accurately predict
responses to stimuli with markedly different statistics (e.g., stimuli
with large changes in mean or variance, or white noise vs.
naturalistic stimuli) \citep{Heitman14}. 

Secondly, GLMs often lack clear
interpretability in terms of underlying mechanisms.  This stands in
contrast to dynamical models designed to capture specific
biophysical variables and processes. In the two-dimensional Izhikevich
model, for example, one variable ($v$) represents the neuron's
membrane potential, and the other variable ($u$) can be understood as
a membrane recovery variable, which reflects K\textsuperscript{+}
channel activation and Na\textsuperscript{+} channel
inactivation.  Despite the fact the GLM filters do not represent specific biophysical variables, in some cases they can still provide insight into underlying biological processes.  For example, recent work provides an interpretation of
the GLM as a synaptic conductance based model with linear
sub-threshold dynamics \citep{Latimer14}.  Work on the SRM has shown that by dividing the effects of spike history into a dynamic threshold and a spike afterpotential, one can in fact measure their separate contributions with intracellular recordings of a neuron's subthreshold voltage; the spike afterpotential can be observed directly in this voltage trace, while the effects on thereshold can be estimated indirectly by noting the absence of firing \citep{Mensi12,Pozzorini13,Gerstner14}.  

A third known limitation of GLMs is that they lack the flexibility to capture some nonlinear response properties of real spike trains.  For example, as point neuron models, GLMs do not reflect the fact that neurons often receive spatially segregated inputs on the dendritic tree, and these inputs can be processed separately and combined nonlinearly \citep{London05}.  Some extensions of the GLM that incorporate nonlinear inputs and multiple subunits \citep{Fitzgerald11,Rajan13,ParkI13gqm,Cui13,McFarland13,Ahrens2008a,Williamson15} may begin to address this issue, but certainly fall short of capturing the full complexity of dendritic processing. For the range of dynamical behaviors considered here, however, we did not find these extensions to be necessary.

For all results shown, we used a GLM with an exponential nonlinearity.
To test the dependence of our results on the form of the nonlinearity,
we also fit GLMs to several of the behaviors with a
``soft-rectifying'' nonlinearity given by $f(x) =
\log\left(1+\exp\left(x\right)\right)$. This function grows only
linearly for large input values, but still has an exponential decay on
its left tail and remains in the family of nonlinearities (convex and
log-concave) for which the GLM log-likelihood is provably concave
\citep{Paninski04}.  For the behaviors tested (tonic spiking, tonic
bursting, phasic spiking, and phasic bursting), our results were
similar to those with an exponential nonlinearity, though generally
not as temporally precise.  This increased precision is likely due to
the fact that an exponential nonlinearity rises more steeply than a
linear-rectifying function, causing the conditional intensity to
accelerate more rapidly from a low-probability to a high-probability of
spiking regime.  
Past studies have found that responses of both retinal ganglion cells and neocortical pyramidal neurons are well described by a GLM with exponential nonlinearity \citep{Pillow08,Jolivet06}.

It is worth noting that for many of the dynamic behaviors studied
here, the GLM parameters were not strongly constrained by the training
data.  (See "Sensitivity to changes in parameter values" in Appendix.)  Slight changes in the model parameters did not produce
noticeable changes in response, at least for the stereotyped range of
input currents and output spike patterns considered.  The filter
parameters were therefore only weakly identifiable, which corresponds
to a likelihood function with a very gradual falloff along certain
directions in parameter space. This uncertainty potentially
complicates interpretation of the filters in terms of functions
performed by the underlying biophysical mechanism. Conversely, it
reveals that the suite of behaviors considered by Izhikevich and
others can be achieved by a range of different GLMs, and that a richer
set of input-output patterns is needed to identify a unique set of GLM
parameters.

\section*{Conclusion}
The GLM has the ability to mimic a wide range of
biophysically realistic behaviors exhibited by real neurons. Although
it is clear there are some forms of nonlinear behavior it cannot
produce, such as frequency-doubled responses of cat Y cells or V1
complex cells to a contrast-reversing grating, our work provides an
existence proof for its ability to exhibit an important range of
response types considered previously only in biophysics and applied
math modeling literature. Moreover, by considering response
stochasticity as another dimension along which real neurons vary, we
have shown that that the GLM can generate response characteristics
ranging from quasi-deterministic to greater-than-Poisson
variability. The GLM therefore provides a flexible yet powerful tool
for studying the dynamics of real neurons and the computations they
carry out. 

\section*{Appendix}

MATLAB code used to generate example responses from Izhikevich neurons and to fit GLMs to these responses is available in a Github repository (\url{https://github.com/aiweber/GLM_and_Izhikevich}).

\subsection*{Izhikevich model simulations}

To generate training data for fitting the GLM, we simulated responses
from an Izhikevich model \citep{Izhikevich03} with parameters set to
published values given for each behavior in \citep{Izhikevich04} (Table \ref{param_table}; parameter values can be found at http://www.\hspace{0pt}izhikevich.\hspace{0pt}org/\hspace{0pt}publications/\hspace{0pt}izhikevich.m).  For each behavior, we generated approximately 20 seconds of
training data using the forward Euler method with fixed time step size
($dt$) given in Table \ref{param_table}.  It should be noted that in some cases,
published parameter values did not produce the desired qualitative
behavior.  In these cases, we tuned the simulation parameters to
achieve the desired behavior.  Parameters marked with an asterisk in
Table 1 indicate those that differ from published values for the
corresponding behavior in \citep{Izhikevich04}.  Additionally, some
behaviors of the Izhikevich neuron are not robust to small changes in
stimulus timing, stimulus amplitude, or time step of integration.  In
particular, we found bistability to be highly dependent on the precise
stimulus timing (onset and duration), stimulus amplitude, and
integration window.  We tuned these values by hand to produce the
desired behavior.

\begin{table}[b!]
\renewcommand{\arraystretch}{1.2}
\vspace{20pt}
\begin{tabular}{ | c | c | c | c | c | c | c |} \hline 
    \textbf{neuron type} & $\boldsymbol{a}$ & $\boldsymbol{b}$ &
                                                                 $\boldsymbol{c}$ & $\boldsymbol{d}$ & $\boldsymbol{I}$ & $\boldsymbol{dt}\, (ms)$ \\ \hline
    tonic spiking & 0.02 & 0.2 & -65 & 6 & 14 & 0.1 \\ \hline
    phasic spiking & 0.02 & 0.25 & -65 & 6 & 0.5 & 0.1 \\ \hline 
    tonic bursting & 0.02 & 0.2 & -50 & 2 & 10* & 0.1\\ \hline 
    phasic bursting & 0.02 & 0.25 & -55 & 0.05 & 0.6 & 0.1\\ \hline 
    mixed mode & 0.02 & 0.2 & -55 & 4 & 10 & 0.1 \\ \hline 
    spike frequency adaptation & 0.01 & 0.2 & -65 & 5* & 20* & 0.1 \\ \hline 
    type I & 0.02 & -0.1 & -55 & 6 & 25 & 1 \\ \hline 
    type II & 0.2 & 0.26 & -65 & 0 & 0.5 & 1 \\ \hline 
    spike latency & 0.02 & 0.2 & -65 & 6 & 3.49* & 0.1 \\ \hline
    resonator & 0.1* & 0.26 & -60 & -1 & 0.3 & 0.5 \\ \hline
    integrator & 0.02 & -0.1 & -66* & 6 & 27.4 & 0.5 \\ \hline
    rebound spike & 0.03 & 0.25 & -60 & 4 & -5 & 0.1 \\ \hline
    rebound burst & 0.03 & 0.25 & -52 & 0 & -5 & 0.1 \\ \hline
    threshold variability & 0.03 & 0.25 & -60 & 4 & 2.3 & 1 \\ \hline
    bistability I & 1 & 1.5 & -60 & 0 & 30* & 0.05 \\ \hline 
    bistability II & 1 & 1.5 & -60 & 0 & 40 & 0.05 \\ \hline 

\end{tabular}
\caption{Parameters of the Izhikevich neuron for dynamic behaviors shown in Figures~2-6,~8-9,~\&~11. Parameters marked with * indicate parameters that differ from those used in \citep{Izhikevich04}. Additionally, only a single form of bistability (bistability I) was presented in \citep{Izhikevich04}.}
\label{param_table}
\end{table}    

\subsection*{GLM fitting and simulations}
A generalized linear model for a single neuron attributes features of a spike train to both stimulus dependence and spike history.  Stimulus dependence is captured by a stimulus filter $k$, and spike-history dependence is captured by a post-spike filter $h$.   $k$ and $h$ are represented with a raised cosine basis to reduce to the dimensionality necessary to fit and ensure smoothness of the filters.  Basis vectors are of the form:
\begin{equation} 
b_j(t) = \frac{1}{2}\cos(a\log[t+c]-\phi_j)+\frac{1}{2}
\end{equation}
for $t$ such that $a\log(t+c) \in [\phi_j-\pi,\phi_j+\pi]$ and 0 elsewhere.  The parameter $c$ determines the extent to which peaks of the basis vectors are linearly spaced, with larger values of $c$ resulting in more linear spacing.  We typically used 6 such basis vectors to fit a 100 ms stimulus filter $k$ and 8 basis vectors to fit a 150 ms post-spike filter $h$, for a total of 15 parameters (including one for $\mu$ that determines baseline firing rate).  In some cases, as few as 7 or as many as 26 parameters were used to fit an individual Izhikevich neuron's behavior.  In general, the fewest number of basis vectors required to reproduce a given behavior were used, though it is likely that by altering specific features of the basis vectors (e.g., their spacing), even fewer parameters would suffice.

We fit the model parameters (weights on the basis functions for $k$,
weights on the basis functions of $h$, and $\mu$) by maximizing the
log-likelihood:
\begin{equation}
\mathcal{L}(\theta) = \sum \limits_{t=spike} \log\lambda(t) - \Delta\sum \limits_{t} \lambda(t)
\end{equation}
where $\Delta$ is the time resolution of $y(t)$.  We used MATLAB's
\texttt{fminunc} function, part of the MATLAB optimization toolbox, to
find the global maximum of the likelihood function.

We simulated the GLM response in time bins of the same size as the corresponding Izhikevich neuron
and computed the single-bin probability of a spike as
 \begin{equation}
 P(y(t) \geq 1 | \lambda(t)) = 1- P(y(t)=0|\lambda(t)) = 1 - \exp(\Delta \lambda(t)),
 \end{equation}
 where $\Delta$ is the time bin size, so that the probability of 0 or
 1 spikes in a bin sums to 1 (resulting in a Bernoulli approximation
 to the Poisson process), disallowing spike counts greater than 1 in a
 single bin.

\subsection*{Systematic variation of filter amplitudes}
In order to more carefully examine the transitions between different behaviors as the stimulus and post-spike filter change, we systematically varied the amplitude of individual filter components and observed the behavior produced.  Each filter was parameterized with 2 components.  The amplitude of one was fixed while the amplitude of the other was varied.  For the stimulus filter, the amplitude of the second component was varied (-1.5 to +0.25).  This allowed us to transition from monophasic to biphasic filters.  The amplitude of the first component was set to be positive (+1), creating an ``ON" filter appropriate for a positive step stimulus.  For the post-spike filter, the amplitude of the first component was varied (-1 to +1).  The amplitude of the second component was set to be negative (-3), ensuring that spiking would be suppressed on longer timescales.  For the post-spike filter we also imposed an absolute refractory period of 5 ms.  Finally, we included a negative baseline drive ($\mu$ = -1) to suppress spontaneous spiking so that the baseline firing rate was zero.

We simulated responses to 25 identical step stimuli for each set of filters and then classified the behaviors as quiescent, phasic spiking, phasic bursting, tonic spiking, or tonic bursting.  The most commonly observed behavior over the 25 repetitions is depicted in Figure~\ref{fig:param_sweep}.  Responses were classified in the following way.  If no spikes were elicited in the first 200 ms of stimulus presentation and fewer than 5 spikes were elicited during the final 10 seconds of stimulus presentation, the behavior was classified as quiescent.  If at least one spike was elicited in the first 200 ms following stimulus onset and fewer than 5 spikes were elicited during the final 10 seconds of stimulus presentation, the behavior was classified as phasic.  Phasic firing patterns were further classified into phasic spiking if only 1 or 2 spikes were elicited in the first 200 ms, and phasic bursting if 3 or more spikes were elicited in the first 200 ms.  The remaining responses were classified as either tonic spiking or tonic bursting in the following manner.  Inter-spike interval distributions were fit with a Gaussian mixture distribution using MATLAB's \texttt{gmdistribution} function.  We fit both a single Gaussian distribution as well as a mixture of two Gaussians and then compared the Akaike information criterion (AIC) values to determine whether the ISI distribution was better fit as a unimodal distribution or a bimodal distribution, with a lower AIC indicating better fit.  If $0.9 \cdot \textrm{AIC}_{\textrm{unimodal}} < \textrm{AIC}_{\textrm{bimodal}}$, the spike train was classified as tonic spiking; otherwise, it was classified as tonic bursting.  (We added the 0.9 factor to create a more stringent standard for what is classified as bursting activity so that the responses that fall into this category are strongly bimodal distributions that would be readily identified as bursting.  Slightly altering the value of this factor, or eliminating it entirely, gives the same qualitative results, but merely shifts the boundary in Figure~\ref{fig:param_sweep} between the tonic spiking and tonic bursting regions.)

 \subsection*{Sensitivity to changes in parameter values}

\begin{figure}[b!]
\vspace{10pt}
\includegraphics[width=1\textwidth]{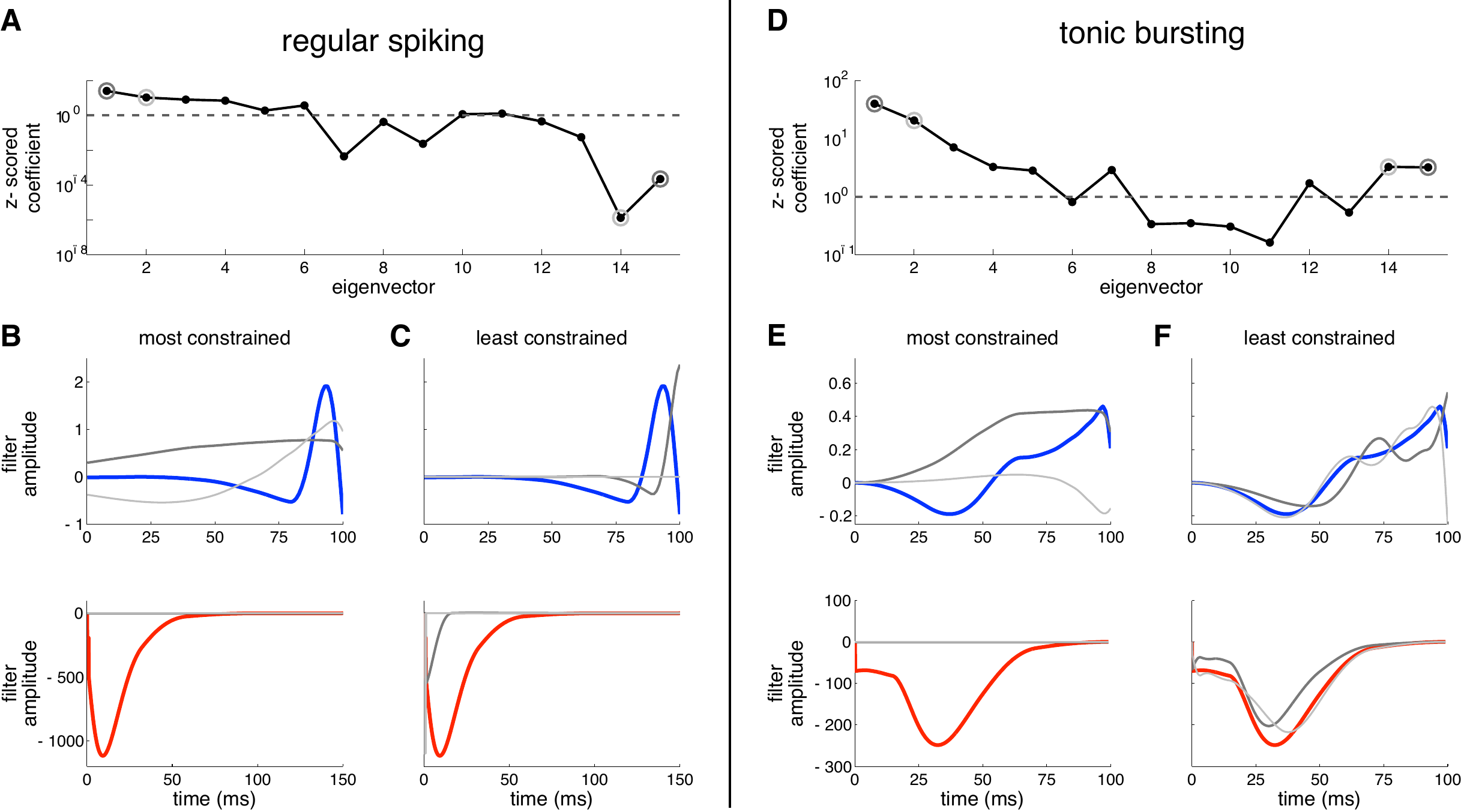}
\caption{{\bf Sensitivity to changes in parameter values for a regular spiking (left) and tonic bursting (right) GLM.} A: Fit coefficient of each eigenvector of Hessian matrix of likelihood, normalized by corresponding eigenvalue. Eigenvectors are in order of decreasing eigenvalues (not necessarily decreasing z-scored eigenvalues).  B: Stimulus filter (top, blue) and spike history filter (bottom, red), along with two most constrained eigenvectors.  These correspond to the largest (dark gray) and second largest (light gray) eigenvalues. Eigenvectors are scaled to size comparable with filters. C: Same as B, for least constrained eigenvectors.  D-F: Same as A-D for tonic bursting neuron.}
\label{fig:sens}
\end{figure}

We wished to investigate how well constrained different features of our fit GLMs were.  To do so, we calculated the eigendecomposition of the Hessian matrix of the likelihood function: $H = Q\Lambda Q^{-1}$.  The Hessian matrix provides a local quadratic approximation to the likelihood function, with eigenvectors $q_i$ pointing along the principal axes and length of these axes proportional to $\frac{1}{\sqrt{\lambda_i}}$.  Thus, larger magnitude eigenvalues indicate greater curvature (i.e., shorter axes) and better constrained directions, while smaller magnitude eigenvalues indicate lower curvature and more poorly constrained directions.  Results of this analysis are shown in Figure~\ref{fig:sens} for both a regular spiking neuron (left) and tonic bursting neuron (right).  For both neurons, the least constrained directions correspond to eigenvectors similar in shape to the best-fit filters or the absolute refractory period.  As such, perturbations in these directions do not result in large changes in the behavior.  Perturbations along eigenvectors corresponding to the most constrained directions, on the other hand, would result in significant changes to the filter shape.  The difference in scale between Panel A and Panel D indicates that overall, parameters for the tonic bursting neuron are more constrained than those for the regular spiking neuron.

\subsection*{Doubly stochastic model with super-Poisson variability}
In Figure~\ref{fig:goris_fig}, we used a negative binomial model to generate
spike trains with greater-than-Poisson variability
\citep{Pillow12,Goris14}.  The negative binomial distribution can be
conceived as a doubly-stochastic model in which the rate of a Poisson
process is modulated by an $iid$ gamma random variable on each
trial. Following \citep{Goris14}, we modeled responses with a
stochastic gain variable $g$ with mean 1 and variance $\sigma_g^2$
that obeys a gamma distribution:
\begin{equation} 
P(g|r,s) = \frac{1}{s^r \Gamma(r)} g^{r-1}\exp \left(-\frac{g}{s}\right),
\end{equation}
where $s = \sigma_g^2$ denotes the scale parameter, $r = 1/\sigma_g^2$
is the shape parameter, and $\Gamma(\cdot)$ represents the gamma
function.

The spike count conditioned on $g$ and a stimulus $S$ for each trial
then obeys a Poisson distribution:
\begin{equation}
P(y|g,S) = \frac{{\left( \Delta g f(S) 
    \right)}^y}{y!}\exp\left(-\Delta g f(S) \right),
\end{equation}
where $\Delta$ is the time bin size, $g$ is the gain, and $f(S)$ is
the tuning curve that specifies the mean response to stimulus $S$.  In
the limit $\sigma_g^2 = 0$, the gain $g$ is deterministically equal to
1 and the spike count is Poisson with mean and variance equal to
$\Delta f(S)$.  For responses with $\sigma_g^2 > 0$, however,
responses are overdispersed relative to the Poisson distribution and
have mean $\Delta f(S)$ and variance $\Delta f(S)( 1+ \sigma_g^2
\Delta f(S))$.

For the results shown in Figure~\ref{fig:goris_fig}, we simulated data from
the negative binomial distribution with a single mean
rate $f(S)$ (100 spikes/s) at three different gain variances $\sigma_g^2 \in
\{.0125, 0.02, 0.05\}$.  Spike counts were drawn $iid$ across trials,
with spike times distributed uniformly within each trial to generate
spike trains suitable for GLM fitting.  We used these spike trains to
fit a GLM to the data associated with each value of $\sigma_g^2$, with
an assumed constant input current for each trial.

\section*{Acknowledgments}
We would like to thank Adrienne Fairhall, Sara Solla, and James Fitzgerald for helpful comments and discussions.


\begin{thebibliography}{}

\bibitem[Ahrens et~al., 2008]{Ahrens2008a}
Ahrens, M.~B., Linden, J.~F., and Sahani, M. (2008).
\newblock Nonlinearities and contextual influences in auditory cortical
  responses modeled with multilinear spectrotemporal methods.
\newblock {\em J Neurosci}, 28(8):1929--1942.

\bibitem[Babadi et~al., 2010]{Babadi2010}
Babadi, B., Casti, A., Xiao, Y., Kaplan, E., and Paninski, L. (2010).
\newblock A generalized linear model of the impact of direct and indirect
  inputs to the lateral geniculate nucleus.
\newblock {\em Journal of Vision}, 10(10):22.

\bibitem[Berry and Meister, 1998]{Berry98}
Berry, M. and Meister, M. (1998).
\newblock Refractoriness and neural precision.
\newblock {\em Journal of Neuroscience}, 18:2200--2211.

\bibitem[Brette and Gerstner, 2005]{Brette05}
Brette, R. and Gerstner, W. (2005).
\newblock Adaptive exponential integrate-and-fire model as an effective
  description of neuronal activity.
\newblock {\em Journal of neurophysiology}, 94(5):3637--3642.

\bibitem[Calabrese et~al., 2011]{Calabrese11}
Calabrese, A., Schumacher, J.~W., Schneider, D.~M., Paninski, L., and Woolley,
  S. M.~N. (2011).
\newblock A generalized linear model for estimating spectrotemporal receptive
  fields from responses to natural sounds.
\newblock {\em PLoS One}, 6(1):e16104.

\bibitem[Cox and Isham, 1980]{Cox80}
Cox, D. and Isham, V. (1980).
\newblock {\em Point Processes}.
\newblock Chapman \& Hall/CRC Monographs on Statistics \& Applied Probability.
  Taylor \& Francis.

\bibitem[Cui et~al., 2013]{Cui13}
Cui, Y., Liu, L.~D., Khawaja, F.~A., Pack, C.~C., and Butts, D.~A. (2013).
\newblock Diverse suppressive influences in area mt and selectivity to complex
  motion features.
\newblock {\em The Journal of Neuroscience}, 33(42):16715--16728.

\bibitem[Fitzgerald et~al., 2011]{Fitzgerald11}
Fitzgerald, J.~D., Rowekamp, R.~J., Sincich, L.~C., and Sharpee, T.~O. (2011).
\newblock Second order dimensionality reduction using minimum and maximum
  mutual information models.
\newblock {\em PLoS Comput Biol}, 7(10):e1002249.

\bibitem[FitzHugh, 1961]{Fitzhugh61}
FitzHugh, R. (1961).
\newblock Impulses and physiological states in theoretical models of nerve
  membrane.
\newblock {\em Biophysical journal}, 1(6):445.

\bibitem[Geisler and Goldberg, 1966]{Geisler66}
Geisler, C.~D. and Goldberg, J.~M. (1966).
\newblock {A Stochastic Model of the Repetitive Activity of Neurons}.
\newblock {\em Biophysical Journal}, 6(1):53--69.

\bibitem[Gerhard et~al., 2017]{Gerhard17}
Gerhard, F., Deger, M., and Truccolo, W. (2017).
\newblock On the stability and dynamics of stochastic spiking neuron models:
  Nonlinear hawkes process and point process glms.
\newblock {\em PLoS computational biology}, 13(2):e1005390.

\bibitem[Gerstner, 1995]{Gerstner95}
Gerstner, W. (1995).
\newblock {Time structure of the activity in neural network models}.
\newblock {\em Physical Review E}, 51(1):738--758.

\bibitem[Gerstner, 2001]{Gerstner01}
Gerstner, W. (2001).
\newblock A framework for spiking neuron models: The spike response model.
\newblock In Moss, F. and Gielen, S., editors, {\em The Handbook of Biological
  Physics}, volume~4, pages 469--516.

\bibitem[Gerstner et~al., 2014]{Gerstner14}
Gerstner, W., Kistler, W.~M., Naud, R., and Paninski, L. (2014).
\newblock {\em Neuronal Dynamics: From Single Neurons to Networks and Models of
  Cognition}.
\newblock Cambridge University Press, New York, NY, USA.

\bibitem[Gerstner and van Hemmen, 1992]{Gerstner92}
Gerstner, W. and van Hemmen, J. (1992).
\newblock {Associative memory in a network of `spiking' neurons}.
\newblock {\em Network: Computation in Neural Systems}, 3(2):139--164.

\bibitem[Gerstner et~al., 1996]{Gerstner96}
Gerstner, W., van Hemmen, J.~L., and Cowan, J.~D. (1996).
\newblock {What matters in neuronal locking?}
\newblock {\em Neural computation}, 8:1653--1676.

\bibitem[Goris et~al., 2014]{Goris14}
Goris, R. L.~T., Movshon, J.~A., and Simoncelli, E.~P. (2014).
\newblock Partitioning neuronal variability.
\newblock {\em Nat Neurosci}, 17(6):858--865.

\bibitem[Heitman et~al., 2014]{Heitman14}
Heitman, A., Greschner, M., Field, G., Li, P., Ahn, D., Sher, A., Litke, A.,
  and Chichilnisky, E. (2014).
\newblock Representation and reconstruction of natural scenes in the primate
  retina.
\newblock In {\em Computational and Systems Neuroscience (CoSyNe) Abstracts},
  pages 134 -- 135.

\bibitem[Hocker and Park, 2017]{Hocker17}
Hocker, D. and Park, I.~M. (2017).
\newblock Multistep inference for generalized linear spiking models curbs
  runaway excitation.
\newblock In {\em 8th International IEEE EMBS Conference On Neural
  Engineering}.

\bibitem[Hodgkin, 1948]{Hodgkin1948}
Hodgkin, A. (1948).
\newblock {The local electric changes associated with repetitive action in a
  non-medullated axon}.
\newblock {\em The Journal of physiology}, 107(2):165--181.

\bibitem[Hodgkin and Huxley, 1952]{HodgkinHuxley52}
Hodgkin, A.~L. and Huxley, A.~F. (1952).
\newblock A quantitative description of membrane current and its application to
  conduction and excitation in nerve.
\newblock {\em J. Physiol.}, 117(4):500--544.

\bibitem[Izhikevich, 2003]{Izhikevich03}
Izhikevich, E.~M. (2003).
\newblock Simple model of spiking neurons.
\newblock {\em IEEE Trans Neural Netw}, 14(6):1569--1572.

\bibitem[Izhikevich, 2004]{Izhikevich04}
Izhikevich, E.~M. (2004).
\newblock Which model to use for cortical spiking neurons?
\newblock {\em IEEE Trans Neural Netw}, 15(5):1063--1070.

\bibitem[Jolivet et~al., 2003]{Jolivet03}
Jolivet, R., Lewis, T., and Gerstner, W. (2003).
\newblock The spike response model: a framework to predict neuronal spike
  trains.
\newblock {\em Springer Lecture notes in computer science}, 2714:846--853.

\bibitem[Jolivet et~al., 2006]{Jolivet06}
Jolivet, R., Rauch, A., L{\"u}scher, H.~R., and Gerstner, W. (2006).
\newblock {Predicting spike timing of neocortical pyramidal neurons by simple
  threshold models}.
\newblock {\em Journal of Computational Neuroscience}, 21(1):35--49.

\bibitem[Keat et~al., 2001]{Keat01}
Keat, J., Reinagel, P., Reid, R., and Meister, M. (2001).
\newblock Predicting every spike: a model for the responses of visual neurons.
\newblock {\em Neuron}, 30:803--817.

\bibitem[Latimer et~al., 2014]{Latimer14}
Latimer, K.~W., Chichilnisky, E.~J., Rieke, F., and Pillow, J.~W. (2014).
\newblock Inferring synaptic conductances from spike trains with a
  biophysically inspired point process model.
\newblock In Ghahramani, Z., Welling, M., Cortes, C., Lawrence, N., and
  Weinberger, K., editors, {\em Advances in Neural Information Processing
  Systems 27}, pages 954--962. Curran Associates, Inc.

\bibitem[London and H{\"{a}}usser, 2005]{London05}
London, M. and H{\"{a}}usser, M. (2005).
\newblock {Dendritic Computation}.
\newblock {\em Annual Review of Neuroscience}, 28(1):503--532.

\bibitem[Mainen and Sejnowski, 1995]{Mainen95}
Mainen, Z. and Sejnowski, T. (1995).
\newblock Reliability of spike timing in neocortical neurons.
\newblock {\em Science}, 268:1503--1506.

\bibitem[McFarland et~al., 2013]{McFarland13}
McFarland, J.~M., Cui, Y., and Butts, D.~A. (2013).
\newblock Inferring nonlinear neuronal computation based on physiologically
  plausible inputs.
\newblock {\em PLoS Comput Biol}, 9(7):e1003143+.

\bibitem[Mensi et~al., 2012]{Mensi12}
Mensi, S., Naud, R., Pozzorini, C., Avermann, M., Petersen, C. C.~H., and
  Gerstner, W. (2012).
\newblock {Parameter extraction and classification of three cortical neuron
  types reveals two distinct adaptation mechanisms}.
\newblock {\em Journal of Neurophysiology}, 107:1756--1775.

\bibitem[Morris and Lecar, 1981]{Morris81}
Morris, C. and Lecar, H. (1981).
\newblock {Voltage oscillations in the barnacle giant muscle fiber.}
\newblock {\em Biophysical journal}, 35(July):193--213.

\bibitem[Nagumo et~al., 1962]{Nagumo62}
Nagumo, J., Arimoto, S., and Yoshizawa, S. (1962).
\newblock {An Active Pulse Transmission Line Simulating Nerve Axon}.
\newblock {\em Proceedings of the IRE}, 117(m V):2061--2070.

\bibitem[Paninski, 2004]{Paninski04}
Paninski, L. (2004).
\newblock Maximum likelihood estimation of cascade point-process neural
  encoding models.
\newblock {\em Network: Computation in Neural Systems}, 15:243--262.

\bibitem[Paninski et~al., 2004]{Paninski04NC}
Paninski, L., Pillow, J.~W., and Simoncelli, E.~P. (2004).
\newblock Maximum likelihood estimation of a stochastic integrate-and-fire
  neural model.
\newblock {\em Neural Computation}, 16:2533--2561.

\bibitem[Park et~al., 2013]{ParkI13gqm}
Park, I.~M., Archer, E.~W., Priebe, N., and Pillow, J.~W. (2013).
\newblock Spectral methods for neural characterization using generalized
  quadratic models.
\newblock In {\em Advances in Neural Information Processing Systems 26}, pages
  2454--2462.

\bibitem[Perkel et~al., 1967]{Perkel67a}
Perkel, D.~H., Gerstein, G.~L., and Moore, G.~P. (1967).
\newblock Neuronal spike trains and stochastic point processes i. the single
  spike train.
\newblock {\em Biophysical Journal}, 7(4):391--418.

\bibitem[Pillow and Scott, 2012]{Pillow12}
Pillow, J. and Scott, J. (2012).
\newblock Fully bayesian inference for neural models with negative-binomial
  spiking.
\newblock In Bartlett, P., Pereira, F., Burges, C., Bottou, L., and Weinberger,
  K., editors, {\em Advances in Neural Information Processing Systems 25},
  pages 1907--1915.

\bibitem[Pillow et~al., 2005]{Pillow05}
Pillow, J.~W., Paninski, L., Uzzell, V.~J., Simoncelli, E.~P., and
  Chichilnisky, E.~J. (2005).
\newblock Prediction and decoding of retinal ganglion cell responses with a
  probabilistic spiking model.
\newblock {\em The Journal of Neuroscience}, 25:11003--11013.

\bibitem[Pillow et~al., 2008]{Pillow08}
Pillow, J.~W., Shlens, J., Paninski, L., Sher, A., Litke, A.~M., and
  Chichilnisky, E. J.~Simoncelli, E.~P. (2008).
\newblock Spatio-temporal correlations and visual signaling in a complete
  neuronal population.
\newblock {\em Nature}, 454:995--999.

\bibitem[Pozzorini et~al., 2013]{Pozzorini13}
Pozzorini, C., Naud, R., Mensi, S., and Gerstner, W. (2013).
\newblock {Temporal whitening by power-law adaptation in neocortical neurons}.
\newblock {\em Nature Neuroscience}, 16(7):942--948.

\bibitem[Rabinowitz et~al., 2015]{Rabinowitz15}
Rabinowitz, N.~C., Goris, R.~L., Cohen, M., and Simoncelli, E. (2015).
\newblock Attention stabilizes the shared gain of v4 populations.
\newblock {\em eLife}.

\bibitem[Rajan et~al., 2013]{Rajan13}
Rajan, K., Marre, O., and Tka{\v c}ik, G. (2013).
\newblock Learning quadratic receptive fields from neural responses to natural
  stimuli.
\newblock {\em Neural Computation}, 25(7):1661--1692.

\bibitem[Schneidman et~al., 1998]{Schneidman98}
Schneidman, E., Freedman, B., and Segev, I. (1998).
\newblock {Ion channel stochasticity may be critical in determining the
  reliability and precision of spike timing.}
\newblock {\em Neural computation}, 10(7):1679--703.

\bibitem[Shadlen and Newsome, 1998]{Shadlen98}
Shadlen, M. and Newsome, W. (1998).
\newblock The variable discharge of cortical neurons: implications for
  connectivity, computation, and information coding.
\newblock {\em Journal of Neuroscience}, 18:3870--3896.

\bibitem[Tolhurst et~al., 1983]{Tolhurst83}
Tolhurst, D.~J., Movshon, J.~A., and Dean, A.~F. (1983).
\newblock The statistical reliability of signals in single neurons in cat and
  monkey visual cortex.
\newblock {\em Vision Res}, 23(8):775--785.

\bibitem[Tomko and Crapper, 1974]{Tomko74}
Tomko, G.~J. and Crapper, D.~R. (1974).
\newblock Neuronal variability: non-stationary responses to identical visual
  stimuli.
\newblock {\em Brain research}, 79(3):405--418.

\bibitem[Truccolo et~al., 2005]{Truccolo05}
Truccolo, W., Eden, U.~T., Fellows, M.~R., Donoghue, J.~P., and Brown, E.~N.
  (2005).
\newblock A point process framework for relating neural spiking activity to
  spiking history, neural ensemble and extrinsic covariate effects.
\newblock {\em J. Neurophysiol}, 93(2):1074--1089.

\bibitem[Uzzell, 2004]{Uzzell04}
Uzzell, V.~J. (2004).
\newblock {Precision of Spike Trains in Primate Retinal Ganglion Cells}.
\newblock {\em Journal of Neurophysiology}, 92(2):780--789.

\bibitem[Weber et~al., 2012]{Weber12}
Weber, F., Machens, C.~K., and Borst, A. (2012).
\newblock Disentangling the functional consequences of the connectivity between
  optic-flow processing neurons.
\newblock {\em Nature neuroscience}, 15(3):441--448.

\bibitem[Weiss, 1966]{Weiss66}
Weiss, T.~F. (1966).
\newblock A model of the peripheral auditory system.
\newblock {\em Kybernetik}, 3(4):153--175.

\bibitem[Williamson et~al., 2015]{Williamson15}
Williamson, R.~S., Sahani, M., and Pillow, J.~W. (2015).
\newblock The equivalence of information-theoretic and likelihood-based methods
  for neural dimensionality reduction.
\newblock {\em PLoS Comput Biol}, 11(4):e1004141.

\bibitem[Zhao and Iyengar, 2010]{Zhao10}
Zhao, M. and Iyengar, S. (2010).
\newblock Nonconvergence in logistic and poisson models for neural spiking.
\newblock {\em Neural Computation}, 22(5):1231--1244.
\newblock PMID: 20100077.

\end{thebibliography}
\end{document}